\documentclass[twocolumn,preprintnumbers,amsmath,amssymb]{revtex4-1}
\usepackage{colortbl}
\usepackage[table]{xcolor}
\usepackage{array}
\usepackage{booktabs}
\usepackage{graphicx}
\usepackage{dcolumn}
\usepackage{bm}

\usepackage{array}
\usepackage{makecell}

\usepackage{tikz}

\definecolor{half}{rgb}{0.91, 0.84, 0.42}
\definecolor{U50to59}{rgb}{0.47,0.8,0.47}
\definecolor{U60to100}{rgb}{0.8,0.15,0.15}
\definecolor{L50to59}{rgb}{0.94,0.5,0.5}
\definecolor{L60to100}{rgb}{0.15,0.55,0.15}

\usepackage{xcolor,colortbl}

\newcommand{\done}{\cellcolor{half}}
\newcommand{\high}{\cellcolor{U50to59}}

\newcommand{\low}{\cellcolor{L50to59}}

\usepackage{ragged2e}

\begin{document}



\title{Anticipating regime shifts in gene expression: The case
  of an autoactivating positive feedback loop}

\author{Yogita Sharma} \affiliation{Department of Mathematics, Indian
  Institute of Technology Ropar, Punjab 140 001, India} \author{Partha
  Sharathi Dutta} \thanks{Corresponding author:
  parthasharathi@iitrpr.ac.in} \affiliation{Department of Mathematics,
  Indian Institute of Technology Ropar, Punjab 140 001, India}
\author{A. K. Gupta} \affiliation{Department of Mathematics, Indian
  Institute of Technology Ropar, Punjab 140 001, India}

\received{:to be included by reviewer}
\date{\today}

\begin{abstract}

Considerable evidence suggests that anticipating sudden shifts from
one state to another in bistable dynamical systems is a challenging
task, examples include ecosystems, financial markets, complex
diseases, etc.  In this paper, we investigate the effects of additive,
multiplicative and cross correlated stochastic perturbations on
determining regime shifts in a bistable gene regulatory system, which
gives rise to two distinct states of low and high concentrations of
protein.  We obtain the stationary probability density and mean first
passage time of the system.  We show that increasing
additive(multiplicative) noise intensity induces regime shift from a
low(high) to a high(low) protein concentration state.  However, an
increase in cross correlation intensity always induces regime shifts
from high to low protein concentration state.  For both bifurcation
(often called tipping point) and noise induced (called stochastic
switching) regime shifts, we further explore the robustness of
recently developed critical slowing down based early warning signal
(EWS) indicators (e.g., rising variance and lag-1 autocorrelation) on
our simulated time series data.  We identify that using EWS
indicators, prediction of an impending bifurcation induced regime
shift is relatively easier than that of a noise induced regime shift
in the considered system.  Moreover, the success of EWS indicators
also strongly depends upon the nature of noise.  Our results establish
the key fact that finding more robust indicator to forewarn regime
shifts for a broader class complex natural systems is still in its
infancy and demands extensive research.

\end{abstract}

\maketitle 

\section{Introduction \label{sec:intro}}

Theoretical as well as experimental studies have indicated that many
complex systems under the influence of stochastic perturbations can
undergo sudden ``regime shifts'' in which they abruptly shift to a
contrasting state.  Such shifts may occur in systems with alternative
stable states \cite{Sc01}.  Well studied examples of regime shifts
include sudden collapse of ecosystems \cite{RiRuKo04}, the onset of
collapse in mutualistic communities \cite{DaBa14}, abrupt climatic
shifts \cite{LeHeKr08}, the crash of markets in global finance
\cite{MaLeSu08}, systemic failures such as the epileptic seizures
\cite{Mc03} and even the eruptive events in spreading fires
\cite{FoWh15}.  Despite rich advances in the theory of complex
systems, understanding the mechanism which trigger the onset of regime
shifts in nature remains a challenge.

There are mainly two types of regime shifts that can occur in systems
with alternative stable states. One is {\em critical transition} which is
associated with the {\em bifurcation points} (so called tipping
points) \cite{Scheffer:2012sc} and another is {\em noise induced
  transition} (also known as stochastic switching)
\cite{Boettiger:2013jc}.  Much effort has been devoted in recent years
in developing {\em early warning signals} of impending regime shifts
between alternative stable states
\citep{CaBr06,CaBr08,GuJa08,ScJo09,Se11,Dakos:2012pone,Kefi:2014pone}.
Such early warning signals can have tremendous impact on managing
natural systems by forewarning the systems under the threat of state
shift, so that appropriate management strategies can be initiated to
prevent a catastrophe.  Recent progress in this direction suggests
that a set of generic statistical indicators (e.g., increase in
variance, autocorrelation) may forewarn an impending transition in a
wide range of complex systems \citep{ScJo09}.  A recent review on
``dynamical disease'' argued that an early detection of regime shifts
even in the field of medical sciences, such as in cardiac arrhythmia's
could be of some help to prevent sudden death \cite{gl15}.  However,
these signals are mostly developed for the phenomenon of critical
slowing down that arises in the vicinity of tipping points
\citep{ScJo09,DaCa15}.  For purely noise induced transitions, such as
we also examine in this paper, there is an active debate about whether
early warning signals can really be useful
\citep{Drake:2013cb,Boettiger:2013jc,DaCa15}.

In genetic regulatory systems, it is known that random cell-to-cell
variations within a genetically identical population can lead to
regime shifts between alternative stable states of gene expression
(i.e., sudden transition in protein concentration)
\citep{KaEl05,MaRaDu07}.  When the underlying genetic system contains
regulatory positive feedback loops, individual cells can exist in
different steady states \citep{AnFeSo04}.  Some cells may, for
example, live in the ``off'' expression state of a particular gene,
whereas others are in the ``on'' expression state \citep{ThPoGr14}.
These stochastic fluctuations in gene expression, commonly referred to
as noise, have been proposed to cause transitions between these states
\citep{Oz04,AcBeOu05}.  A well known example of bistable gene
expression with coupled stochastic transition is the induction of the
{\em lac} operon in {\em E. coli} which results in the synthesis of
protein {\em $\beta$-galactosidase} required for breaking up sugar
molecules and releasing energy to the cell \cite{Muhi}.  Experimental
study on {\em $\beta$-galactosidase} shows that sudden transition from
unregulated (low level) to regulated (high level) {\em
  $\beta$-galactosidase} state of lac operon occurs at a critical
point of an inducer concentration \cite{Oz04}. A potential application
of early warning signals in gene expression is to identify increased
risk of sudden transitions in protein concentration and prevent
complex disease onset \cite{gl15,TrAn15,ChLi12}.

Recently the concept of regime shifts with associated early warning
signals has been used in systems medicine \cite{TrAn15}. The
expectation is to foresee a sudden catastrophic shift in health
condition which may result in a extreme transition to a disease state.
Now a days it is believed that a detailed understanding of regime
shifts in disease onset will provide broad applications in the field
of medicine \cite{TrAn15}.  Already detected sudden transitions in
medical science include epileptic seizures \cite{McSmTa:2003NatMed},
depression \cite{Leemput07012014}, pulmonary disease
\cite{Venegas:2005Nat}, diabetes mellitus \cite{LiLiZh2013}, etc.  For
the case of type 1 diabetes mellitus, {\em $\beta$-cells} in the
pancreas do not produce protein hormone {\em insulin}, which sometimes
is a consequence of ``switched off'' state of HLA-encoding genes in
the cell (``switched on'' state of genes in $\beta$-cell corresponds
to activation of insulin production).  This is due to the fact that
genes carry the instructions that cells use for protein production.
Hence, bistability via ``switching on'' or ``off'' states in ``gene
expression'' is also an important topic to study for regime shifts in
systems medicine \cite{TrAn15}.

In this paper, we study a stochastic version of bistable gene
regulatory positive feedback loop model \cite{Ch08} to explore the
robustness of early warning indicators of regime shifts, for both
cases, the critical transition and noise induced transition.  We
investigate the effect of additive and multiplicative noise
intensities, and cross correlation intensity between two noises
 on the model by calculating the probability density and 
potential function.We find that increasing the intensity of additive noise induces regime
shifts from low to high protein concentration state and vice versa for
an increase in multiplicative noise intensity.  Whereas an increase in
the cross correlation intensity from a negative to a positive value
between the two noises induce regime shifts from high to low protein
concentration state. We also compute mean first passage time (MFPT)
for escape over the potential barrier. We discuss how one can regulate
the production levels of protein. To this end, we apply the early
warning signals of regime shifts \citep{ScJo09} on the simulated time
series data of the stochastic model to examine how successfully one
can forewarn regime shifts.  Our work presents a novel framework for
using early warning signals to identify regime shifts, and also their
key limitations, in a gene regulatory system.  Finally, in the
discussion section, we conclude the paper with a discussion of the
main results together with the applicability and importance of our
study.


\section{A model of genetic regulation \label{sec2}}

\subsection{Deterministic description}

\begin{figure}[h!]
\centering
\includegraphics[width=0.96\columnwidth]{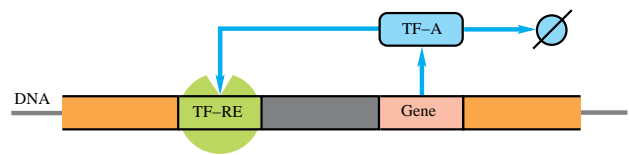}
\caption{\label{sch} A schematic picture of genetic expression for an
  auto-activating positive feedback loop.  The expression of gene
  leads to protein (TF-A) that after oligomerization binds to its own
  promoter (TF-RE), acting as an self activator. Degradation of the
  protein is denoted by the slashed circle ($\varnothing$).}
\end{figure}

It is well known that autoactivating positive feedback loop is one
of the simplest circuit motifs able to exhibit bistable states in gene
regulatory process \cite{BeSeSe01,TyChNo03,LiJi04,Ch08,Frcasaib12}.  In the
circuit, a single gene encodes a transcriptional factor activator
(TF-A), and the TF-A activates its own transcription when bind to a
responsive element (TF-RE) in the DNA sequence (see Fig.~\ref{sch} for
a schematic representation).  The transcription factor activator TF-A
is referred to as regulatory protein, is used to control genetic
regulation and acts as a pathway mediating a cellular response to a
stimulus.  The TF-A constructs a homodimer which then binds to TF-RE
or DNA regulatory site.  Gene incorporates a TF-RE and when homodimers
bind to the TF-RE, transcription of TF-A is increased.  Homodimer
binding to responsive element is independent phosphorylation of
dimers.  However, transcription can be activated only by
phosphorylated TF-A dimers. Some dimers phosphorylation will depend on
activity of kinases and phosphatases which is controlled by external
signals.  Hence, this genetic circuit incorporates signal-activated
transcription as well as positive feedback on TF-A synthesis.  Let,
$X$ denotes the protein TF-A, $D$($D^*$) denotes the unbound(bound)
state of the DNA promoter, then the equilibrium reactions can be
written as:
\begin{center}
$\displaystyle \varnothing
  \underset{k_{deg}}{\stackrel{RV}{\rightleftharpoons}} X,$\\ $D+nX$
  $\displaystyle
  \underset{k_{2}}{\stackrel{k_{1}/V^{n}}{\rightleftharpoons}}$
  $D^{*},$\\ $D^{*}$ $\displaystyle \overset{k_{p}}\rightarrow$
  $D^{*}+X,$
\end{center}
where $R$ is the basal expression rate, $V$ is cellular volume, $X$ is
degraded with a rate constant $k_{deg}$, $n$ represents cooperativity
in binding, binding(unbinding) of TF-A homodimer to DNA regulatory
site with a rate constant $k_1$ ($k_2$) and $k_p$ is protein
production rate of TF-A.  Letting $x=[X]$, $d=[D]$ and $d^*=[D^*]$,
the rate equation of the evolution of concentration of TF-A can be
written as \cite{WeBu13}:
\begin{equation}\label{Eq:1}
\frac{dx}{dt} = R+a\frac{x^n}{K_{d}+x^n}-k_{deg}x,
\end{equation}
where $a=k_p(d+d^*)$ is the maximum production(i.e., transcription)
rate, $K_{d}=k_2/k_1$ is the dissociation constant of TF-A from
TF-RE. A more detail derivation of Eq.~(\ref{Eq:1}) is given in
\cite{WeBu13}.

\begin{figure}[h!]
\centering
\includegraphics[width=0.96\columnwidth]{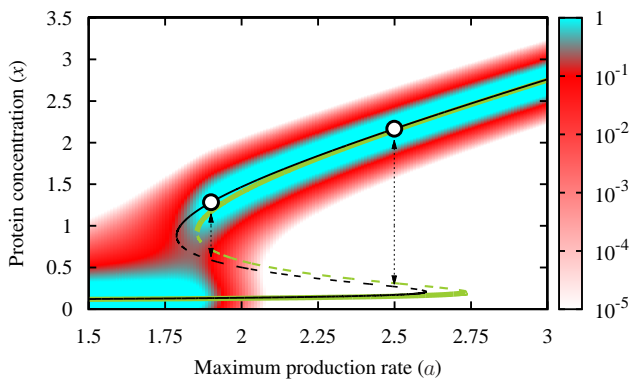} 
\caption{\label{Fig:bif_PD} Bifurcation diagram for the deterministic
  model (black) and the stochastic model (green). The parameter values
  are $r=0.1$ (for the deterministic model) and $r=0.1$ with
  $\sigma_1=0.1$, $\sigma_2=0.1$ and $\lambda=0.1$ (for the stochastic
  model). Stable steady states are marked with continuous lines and
  unstable steady states are marked with dashed lines, respectively.
  The stationary probability distribution for the additive noise
  model, Eq.~(\ref{Eq:3} ), for different values of the production
  rate $a$ is shown in cyan-red-white scale (color bar, logarithmic
  scale).}
\end{figure}

We can rewrite the dimensionless version of Eq.~(\ref{Eq:1}) as
\begin{equation}\label{Eq:diml}
\frac{d\tilde{x}}{d\tilde{t}}=\tilde{r}+\tilde{a} \frac{\tilde{x}^n}
{1+\tilde{x}^n} -\tilde{x},
\end{equation}
where $\tilde{x}= \frac{x}{\sqrt[n]{K_{d}}}$, $\tilde{t}=k_{deg}t$,
$\tilde{a}=\frac{a}{k_{deg}\sqrt[n]{K_{d}}}$, and
$\tilde{r}=\frac{R}{k_{deg}\sqrt[n]{K_{d}}}$.  The analyses in this
work have been carried out using the above model with dimensionless
variable and parameters.  For certain parameter values,
Eq.~(\ref{Eq:diml}) has three equilibria, of which two are attractors
and third is a saddle point intermediate between the two attractors
(see Fig.~\ref{Fig:bif_PD}).  The saddle point works as a basin
boundary between the two stable equilibrium points.  A thorough
analysis of the deterministic model is given in \cite{Sm98}.
Hereafter, in Eq.~(\ref{Eq:diml}) for sake of simplicity we use $x$,
$t$, $r$ and $a$ in place of $\tilde x$, $\tilde t$, $\tilde r$ and
$\tilde a$ respectively, and Eq.~(\ref{Eq:diml}) becomes:
\begin{equation}\label{Eq:Fl}
\frac{dx}{dt}=r+a \frac{x^n}
{1+x^n} -x\;.
\end{equation}
If $n>1$ and $0 < r < 1/3\sqrt{3}$, then Eq.~(\ref{Eq:Fl}) exhibits
bistability for a range of the parameter $a$ \cite{WeBu13}.  To avoid lengthy
calculations, we now replace $n=2$ in Eq.~(\ref{Eq:Fl}), in the rest of
this paper.
To model (\ref{Eq:Fl}), we investigate the effects of additive and
multiplicative noise on the alternative steady states of the gene
regulatory circuit.  We make this attempt because in the presence of
noise, a bistable system may trigger {\em regime shifts}.

\subsection{Stochastic description}

It is well known that noise is inherent in any natural system.  In
this work, we consider that dynamics of the protein concentration ($x$)
are affected by {\em multiplicative} and {\em additive} noises,
similar to Hasty et al. \cite{Ha00}.  These noise terms can induce
sudden shifts in the concentration of protein ($x$).  In a bistable
system, in the absence of noise the system will eventually converge to
one of its two stable fixed points.  In which fixed point it will
converge depends upon the initial condition.  However, the
presence of noise in the system will cause fluctuations in the steady
states \cite{WhoRle84}, which may lead to switching between two
different stable states \cite{DaCa15} or there can be sudden
transition from one stable state to the other stable state
\cite{DaCa15}.  In order to introduce the multiplicative and additive
noise terms in Eq.~(\ref{Eq:Fl}), we consider the one-variable Langevin
equation in the general form as:
\begin{equation}\label{Eq:3}
\frac{dx}{dt}=f(x)+g(x)\xi (t)+\eta (t) ,
\end{equation} 
where $\xi (t)$ and $\eta (t)$ represent Gaussian white noise.  These
noise terms have the following statistical properties:
$\langle \xi (t) \rangle  =  \langle \eta (t) \rangle = 0$, 
$\langle \xi (t) \xi (t')\rangle  =  2\sigma_1\delta(t-t')$, 
$\langle \eta (t) \eta (t')\rangle  =  2\sigma_2\delta(t-t')$, and 
$\langle \xi (t) \eta (t')\rangle  =  \langle
\eta (t) \xi (t') \rangle = 2\lambda \sqrt{\sigma_1\sigma_2}\;\delta(t-t')$,
where $\sigma_{1}$ and $\sigma_{2}$ measure the level of noise
strengths of $\xi (t)$ and $\eta (t)$ respectively, $\lambda$ is the
cross correlation between them, and $t$ and $t'$ denote two different
moments. 

\begin{figure}[!ht]
\begin{center}
{\footnotesize (a)} $~~~~~~~~~~~~~~~~~~~~~~~~~~~~~~~~~~~~~~~~~~~~~~~~~~~~~~~$\\
\vspace{-0.13in}
\resizebox{!}{2.02in}{\includegraphics{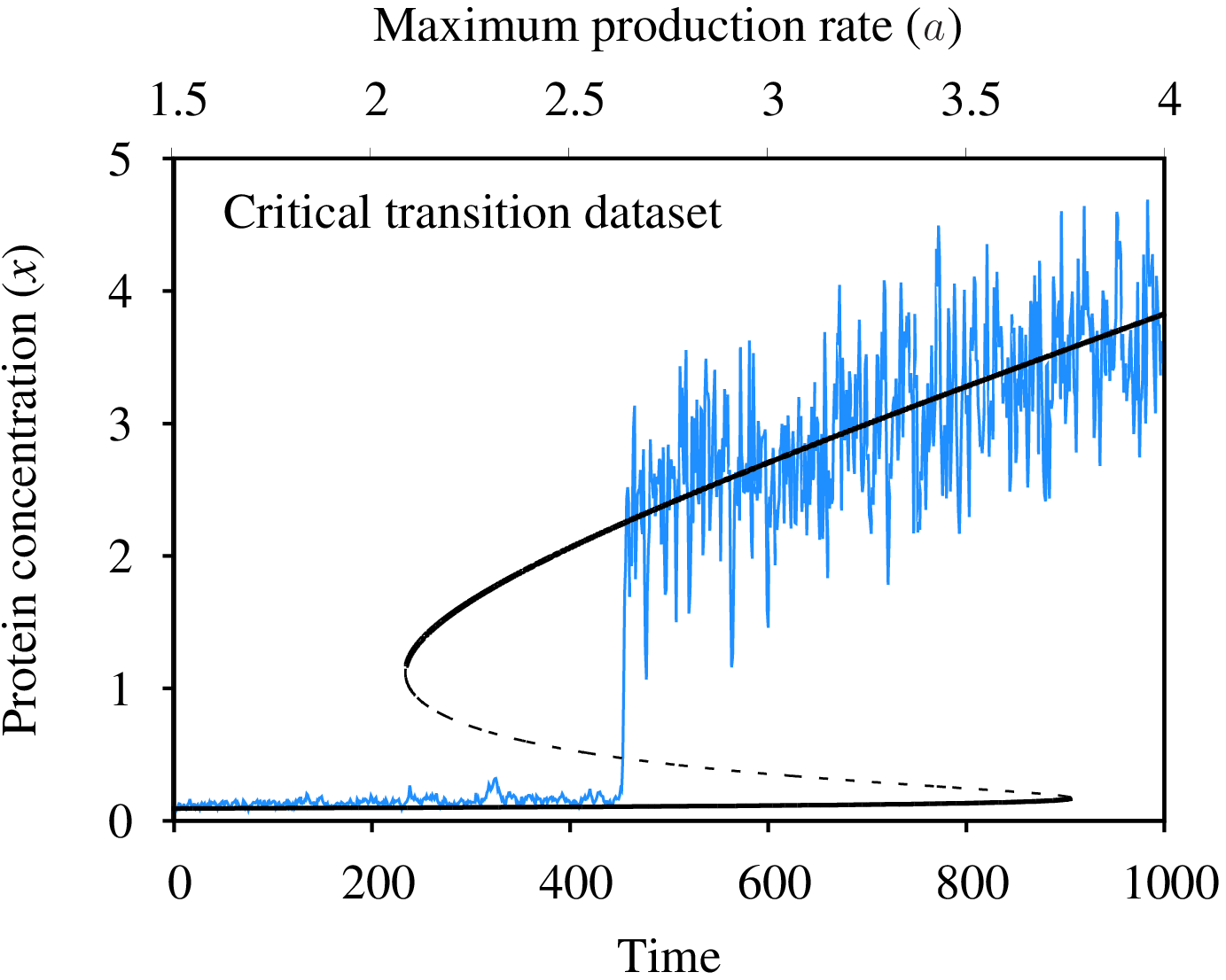}}\\
{\footnotesize (b)} $~~~~~~~~~~~~~~~~~~~~~~~~~~~~~~~~~~~~~~~~~~~~~~~~~~~~~~~$\\
\vspace{-0.08in}
\resizebox{!}{1.74in}{\includegraphics{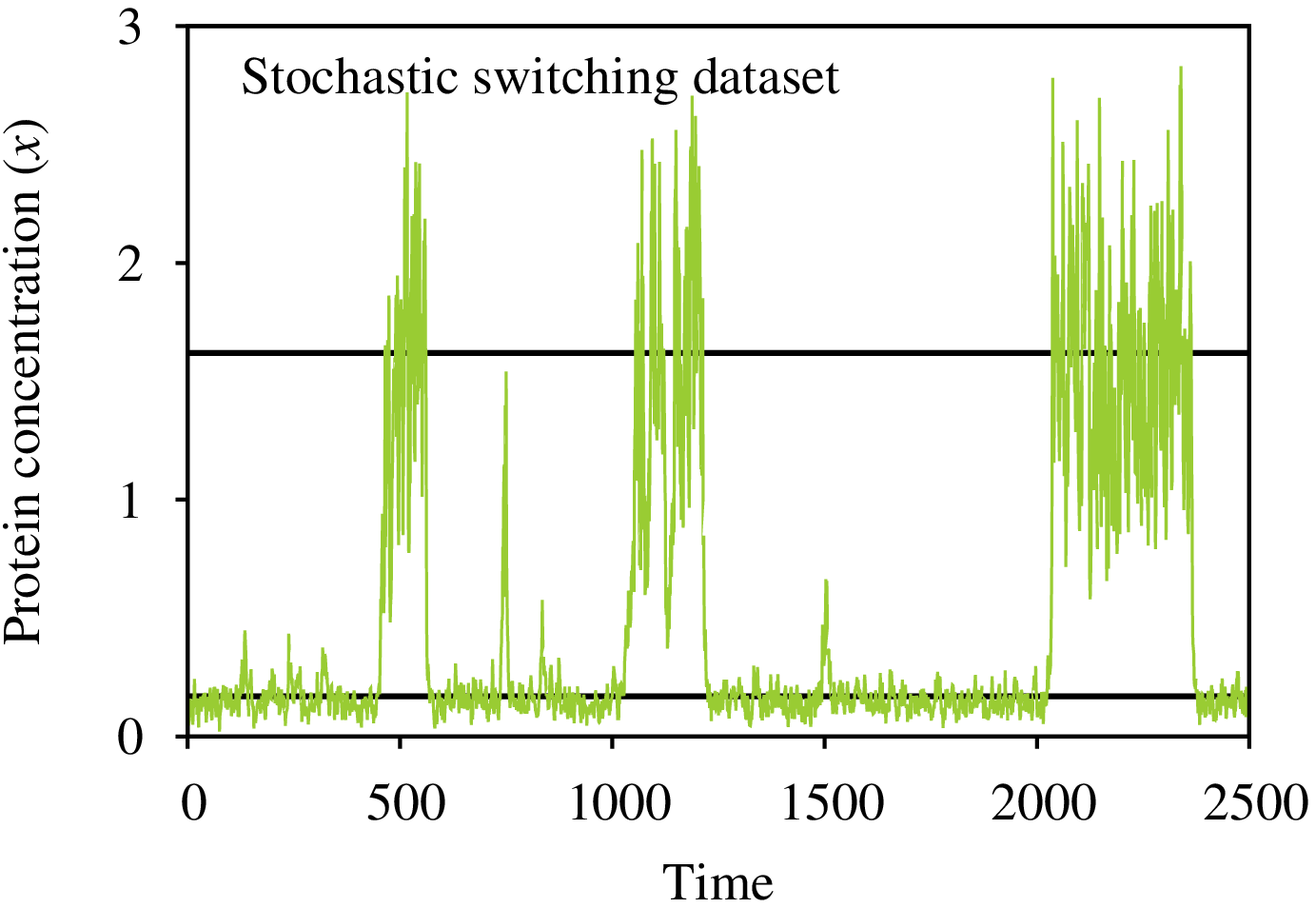}}
\caption{\label{STS} Example time series datasets exhibiting regime
  shifts: (a) Critical transition (associated with bifurcation) and
  (b) Stochastic switching (purely noise induced) for the considered
  stochastic model. See text for more details.}
\end{center}
\end{figure}

In Eq.~(\ref{Eq:3}) the additive noise $\eta (t)$ alters the
background protein production \cite{Ha00}.  It is also known that in
gene expression, transcription is a complex sequence of reactions
\cite{Leh_book}, thus it is expected that this part of the gene
regulatory sequence is also to be affected by fluctuations of many
intrinsic or extrinsic parameters. This implies the fact that the
transcription rate ($a$) can be considered as a random variable
\cite{Ha00}.  To vary the transcription rate stochastically, we
consider $a \rightarrow a+\xi (t)$.  Hence, with the aforementioned
modification, Eq.~(\ref{Eq:Fl}) (for $n=2$) becomes:
\begin{eqnarray}\label{Eq:main}
\frac{dx}{dt} & = & r+\frac{ax^2}{1+x^2}-x+\frac{x^2}{1+x^2}\xi(t)+\eta
(t)\;, \\
& = & f(x)+g(x)\xi (t)+\eta (t)\;\nonumber,
\end{eqnarray}
where, $f(x) = r+\frac{a x^2}{1+x^2} - x$ and $g(x) =
\frac{x^2}{1+x^2}$.  Hence, the noise $\xi (t)$ is multiplicative, as
compared to the additive noise $\eta (t)$.

In the next section we study the stochastic model Eq.~(\ref{Eq:main})
through a combination of {\em analytical} and {\em simulation}
techniques.  Our main aim is to investigate the effects of
multiplicative noise intensity $\sigma_1$, additive noise intensity
$\sigma_2$ and cross correlation strength $\lambda$ between two noises
on the regime shifts between the high and low protein concentration
states.  In the analytical technique these effects are studied by
calculating {\em probability densities, potential functions} and {\em
  MFPT}. The simulation technique is complementary to the analytical
technique, showing how the dynamical properties are captured in our
analytical results can be seen within individual realizations based on
example parameter sets, and adding information about how observed
protein concentrations are arranged in time in these examples.  In
simulations, we also produce {\em time series} exhibiting regime
shifts that can be analyzed using the same techniques as could be
applied to {\em real time series data} (see Figs.~\ref{STS}(a)-(b)).
The example times series can be divided into two broader classes: (a)
critical transition time series (Fig.~\ref{STS}(a)) and (b) purely
noise induced transition time series (Fig.~\ref{STS}(b)).

\section{Results}

\subsection{Fokker-Planck equation and stationary probability density
function \label{sec3}}

We begin this section by writing down the Fokker-Planck equation for
the evolution of probability density of the dynamical variable $x$
\cite{WhoRle84}.  Let $P(x,t)$ denotes the probability density, which
is the probability that the protein concentration attains the value
$x$ at time $t$. Then, the Fokker-Planck equation (FPE) of $P(x,t)$
corresponding to Eq.~(\ref{Eq:main}) is given by \cite{Ri96}:
\begin{equation}\label{Eq:FP1}
\frac{\partial P(x,t)}{\partial t}=-\frac{\partial}{\partial x}[A(x)P(x,t)]
+\frac{\partial^2}{\partial x^2}[B(x)P(x,t)]\;, 
\end{equation}
where
\begin{eqnarray*} 
A(x) & = & f(x)+\sigma_1g(x)g'(x)+\lambda\sqrt{\sigma_1\sigma_2}\;g'(x), \;\text{and}\\ 
B(x) & = & \sigma_1\left[g(x)\right]^2 + \sigma_2
+2\lambda\sqrt{\sigma_1\sigma_2}\; g(x).
\end{eqnarray*}
The limit of $P(x,t)$ as $t \rightarrow \infty$ yields the {\em
  stationary} probability density function (SPDF) of $x$, which we
denote as $P_{s}(x)$.  The SPDF $P_{s}(x)$, which is the stationary
solution of the FPE in Eq.~(\ref{Eq:FP1}) is given by \cite{Ri96} (see
Appendix S1 for more details) :
\begin{eqnarray}
P_{s}(x) & = &
\frac{N_{c}}{B(x)}\exp\left[\int^x\frac{A(x')}{B(x')}dx'\right],\nonumber
\\ & = & \frac{N_{c}}{\sigma_1\left[g(x)\right]^2 +
  \sigma_2+2\lambda\sqrt{\sigma_1\sigma_2}\; g(x)}\nonumber \times
\\ && \!\!\!\!\!\!\!\!\!
\exp\left[\int^x\frac{f(x')+\sigma_1g(x')g'(x')+\lambda\sqrt{\sigma_1\sigma_2}\;
    g'(x')}{\sigma_1\left[g(x')\right]^2 +
    \sigma_2+2\lambda\sqrt{\sigma_1\sigma_2}\;
    g(x')}dx'\right],\nonumber\\
\label{Eq:SPDF1}
\end{eqnarray}
\begin{figure}
\centering
\includegraphics[width=0.85\columnwidth]{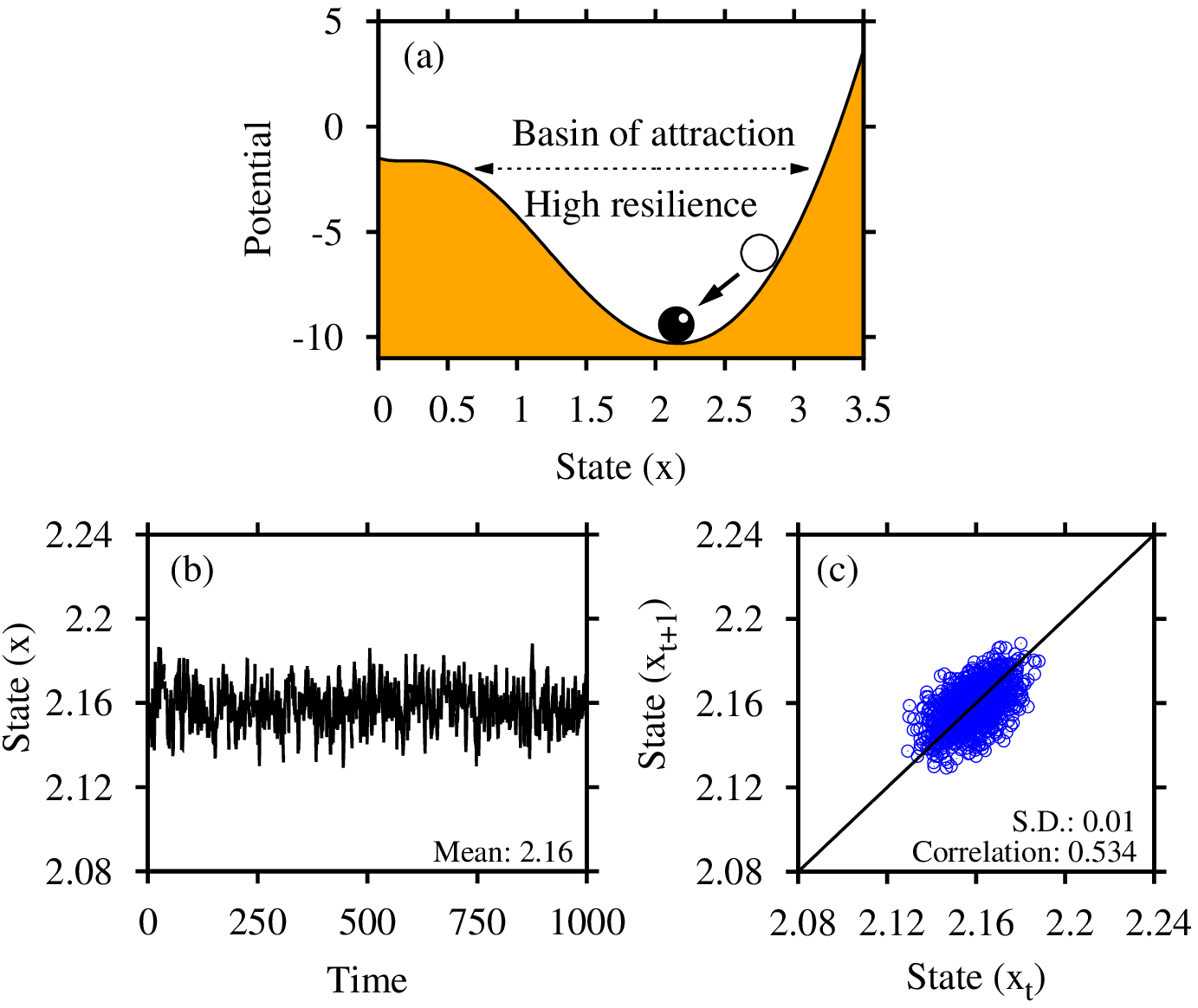}\\
\includegraphics[width=0.85\columnwidth]{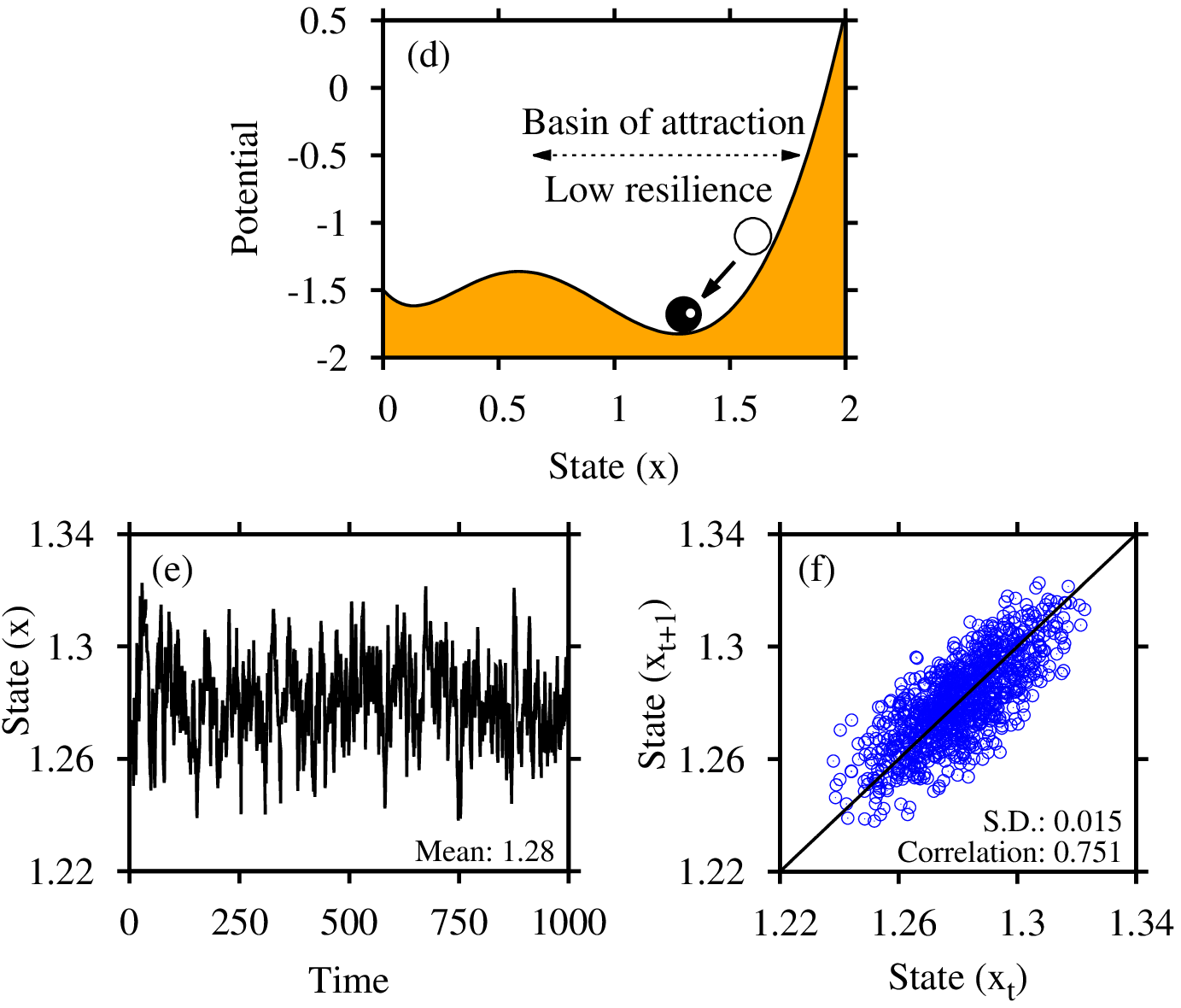}
\caption{\label{Fig:resili} Potential landscapes demonstrating how
  changes in a system parameter can cause decrease in resilience of
  equilibrium point.  Recovery rates upon stochastic fluctuations are
  lower if the basin of attraction is small (d) than that of a larger
  basin of attraction (a).  The effect of reduced resilience can be
  determined by stochastic fluctuations induced in a system state ((b)
  and (e)) as increased standard deviation (S.D.)  and lag-1
  autocorrelation ((c) and (f)).  Data sets to plot this figure
  are generated from Eq.~(\ref{Eq:main}) with $r=0.1$, $\sigma_1=0.0$,
  $\sigma_2=0.05$: (a-c) $a=2.5$ and (d-f) $a=1.9$.}
\end{figure}
where $N_c$ is normalization constant.  Equation~(\ref{Eq:SPDF1}) can
also be put in the form:
\begin{equation}\label{Eq:SPDFPot1}
\displaystyle P_{s}(x)=N_ce^{-\phi(x)}\;,
\end{equation}
where
\begin{eqnarray} 
\phi(x)=\frac{1}{2}\ln \left[\sigma_{1}\left[g(x)\right]^{2} +
  \sigma_2+ 2\lambda\sqrt{\sigma_{1}\sigma_{2}}\;g(x)\right]
\nonumber\\ -\int^{x}\frac{f(x')dx'}{\sigma_{1}\left[g(x')\right]^2
 +\sigma_{2} +2\lambda\sqrt{\sigma_{1}\sigma_{2}}\;g(x')}\;,\label{Eq:POT1}
\end{eqnarray}
is called stochastic potential of the system \cite{Ri96}.  The
potential function maps the equilibria of dynamical systems and their
basins of attraction, by analogy to a ``energy landscape'' in which
the system state tends to move ``downhill''.  Extending this concept
to stochastic dynamical systems gives a probabilistic potential
$\phi(x)$ that complements the SPDF $P_{s}(x)$ in characterizing the
asymptotic behavior of the considered system.  Next we calculate
$\phi(x)$ and $P_{s}(x)$ for three different cases concerning the
effects of multiplicative and additive noise on anticipating regime
shifts in gene expression.  We also calculate bifurcation diagram (see
Fig.~\ref{Fig:bif_PD}) with changing the maximum transcription rate
$a$ for both the deterministic and stochastic model.  Note that, there
is an enlargement of the bistability region for the case of correlated
noise.  In Fig.~\ref{Fig:bif_PD}, we depict the stationary probability
distribution of the additive noise model for different values of $a$,
which is shown in white-red-cyan colorbar.  This gives an idea about
how extrema of stationary probability distribution is changing with
the parameter $a$.  Now, using analytical techniques we first
determine the parameter space where the system persists bistability
still in the presence of stochasticity.  Then, for a specific set of
parameter values, we simulate time series of the system and finally
using EWS indicators \cite{Scheffer:2012sc}, we will try to detect
forthcoming regime shifts.


\subsubsection{Additive noise} \label{SS:ANoi}

First, we present the effect of an additive external noise source on
the regime shifts between high and low concentrations level of protein
in gene expression.  Hence, only the additive noise term is present in
Eq.~(\ref{Eq:main}) and we assume that $\sigma_{1}=0$. Bifurcation
diagram and stationary solutions of the additive noise model are same
as for the deterministic model \cite{Frcasaib12}.  For $\sigma_{1}=0$,
the FPE (\ref{Eq:FP1}) can be written as:
\begin{equation}\label{Eq:padd}
\frac{\partial P(x,t)}{\partial t}=-\frac{\partial}{\partial
  x}[f(x)P(x,t)]+\sigma_{2}\frac{\partial^2}{\partial x^2}[P(x,t)],
\end{equation}
and similar to Eq.~(\ref{Eq:SPDFPot1}) the stationary solution
$P_s(x)$ is written as:
\begin{equation}
P_{s}(x)=N_Ae^{-\phi(x)}\;,
\end{equation}
where 
\begin{equation}\label{Eq:StocPotAN}
\phi(x)=\frac{1}{2}\ln\sigma_{2}-\frac{1}{\sigma_{2}}\int^{x} f(x')dx'\;,
\end{equation}
is the potential and $N_A$ is the normalization constant.

\begin{figure}
\begin{center}
\begin{tabular}{ll}
\resizebox{!}{1.1in}{\includegraphics{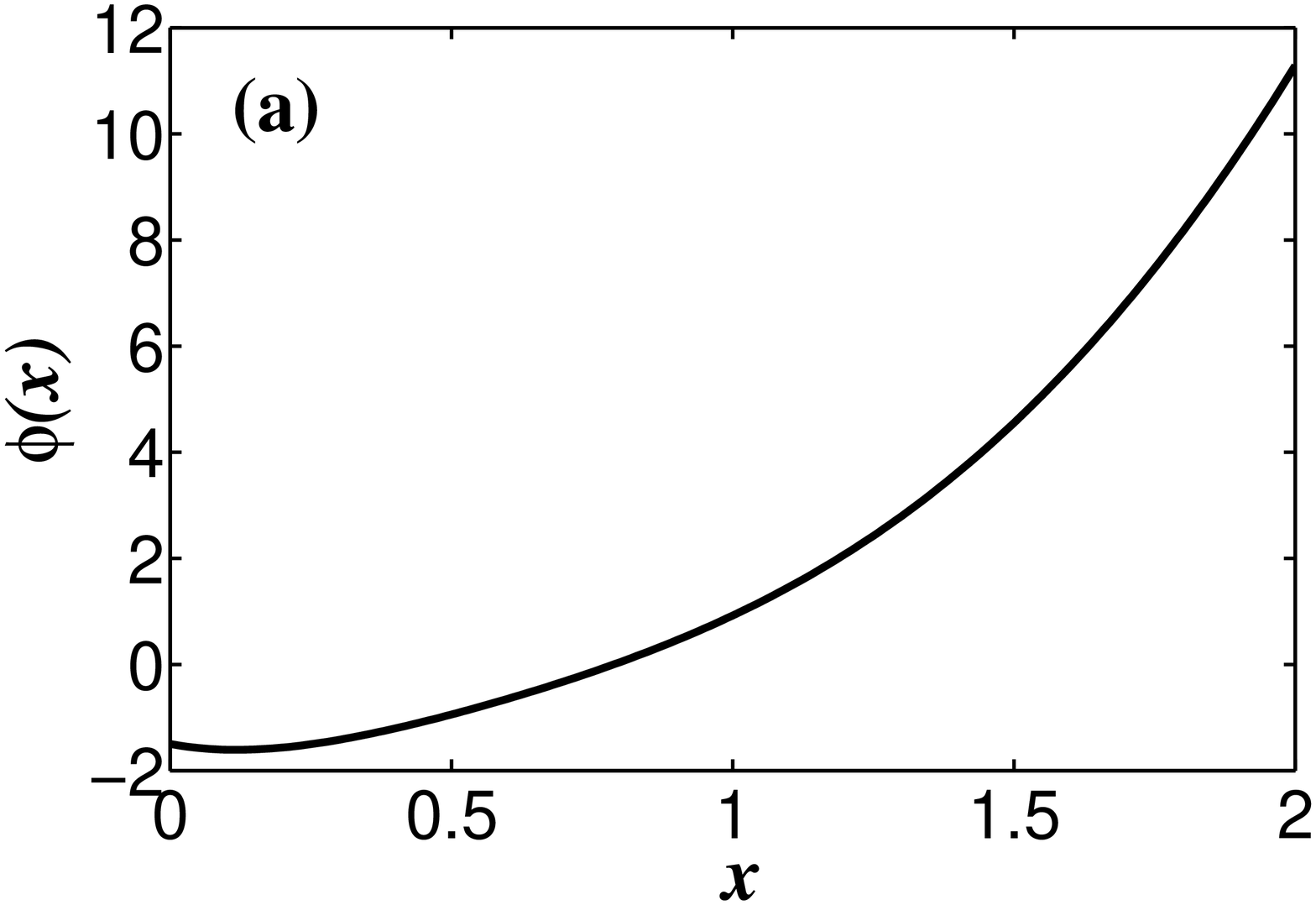}} & \hspace{-.25in}
\resizebox{!}{1.12in}{\includegraphics{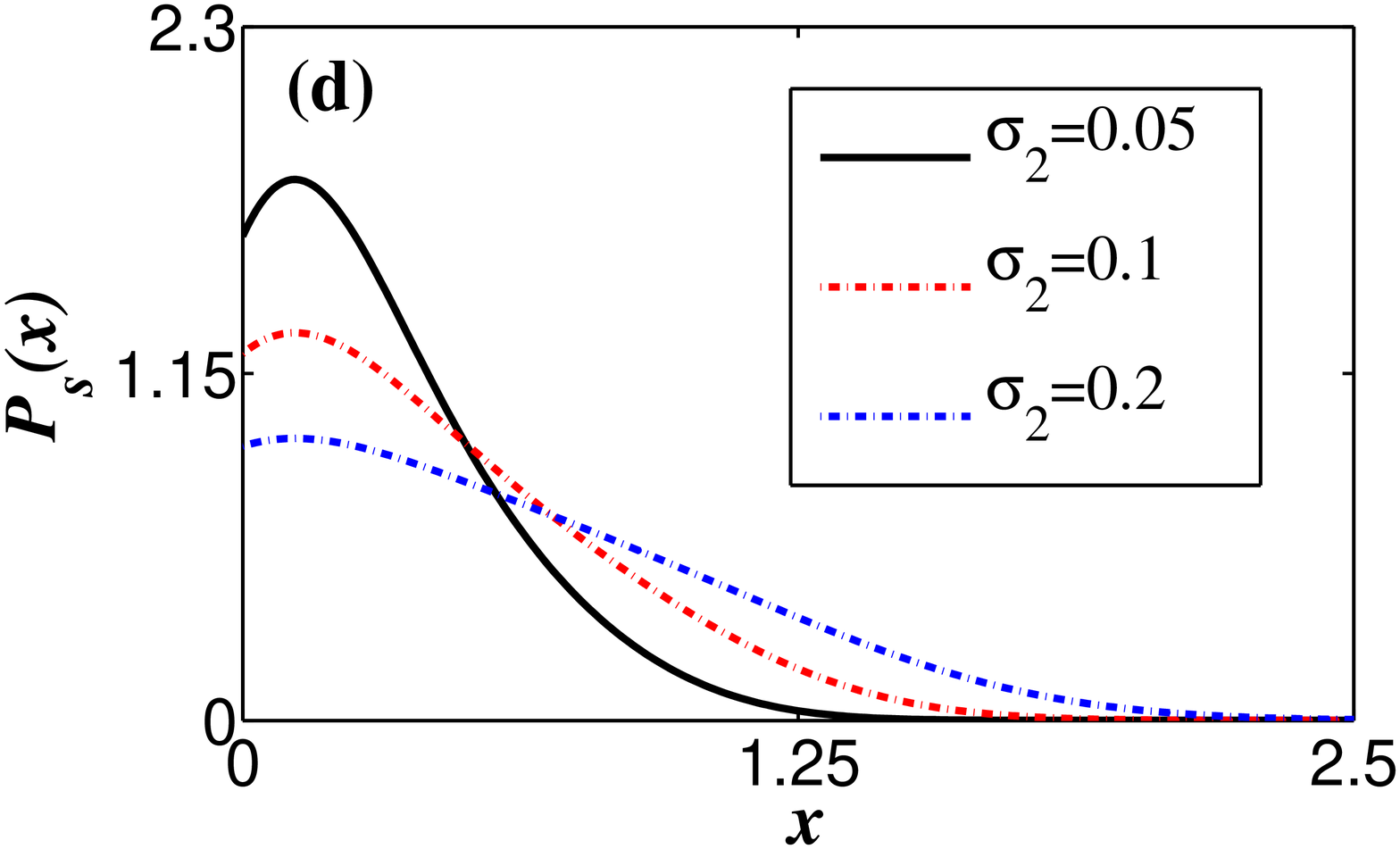}} \\
\resizebox{!}{1.09in}{\includegraphics{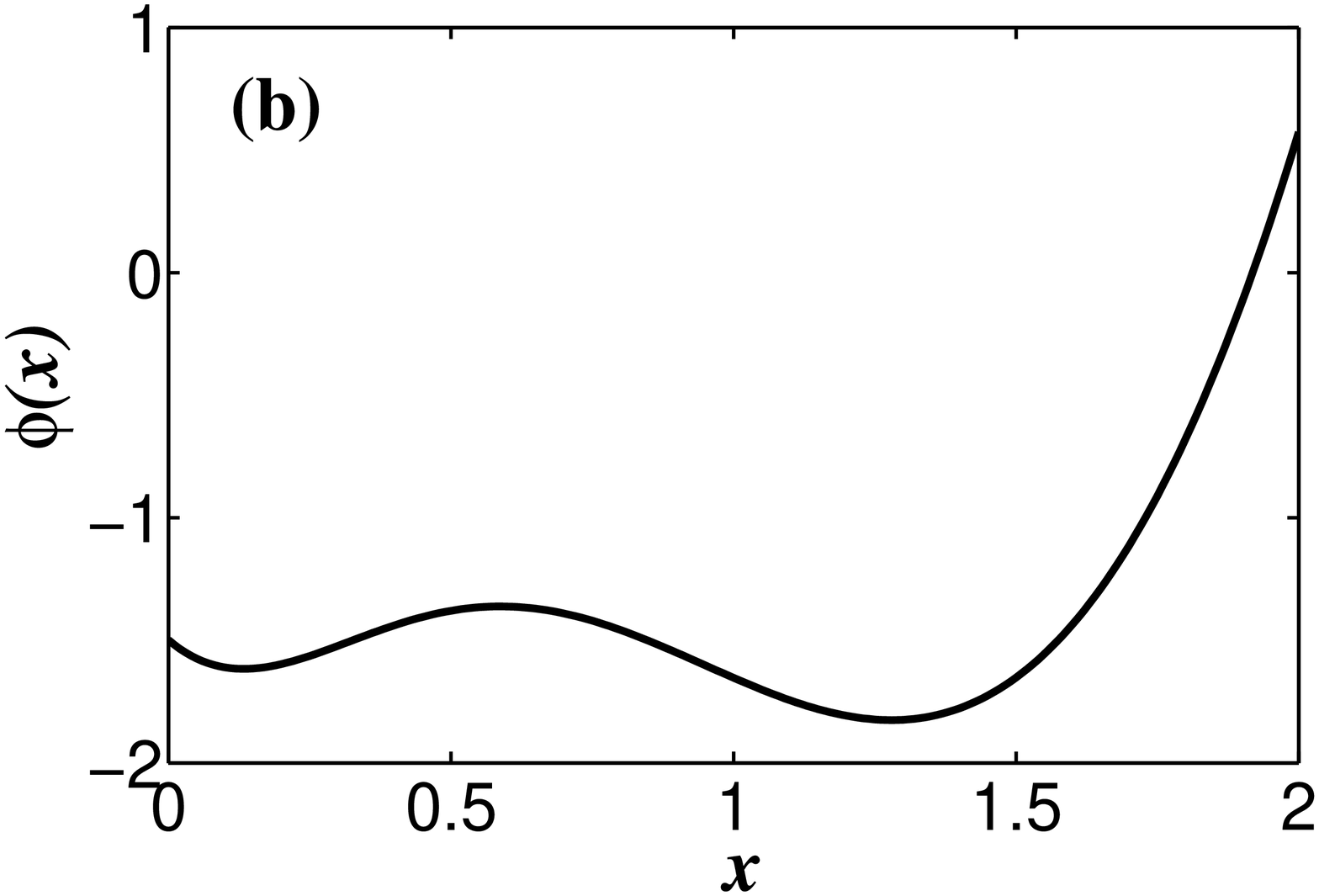}} & \hspace{-.2in}
\resizebox{!}{1.1in}{\includegraphics{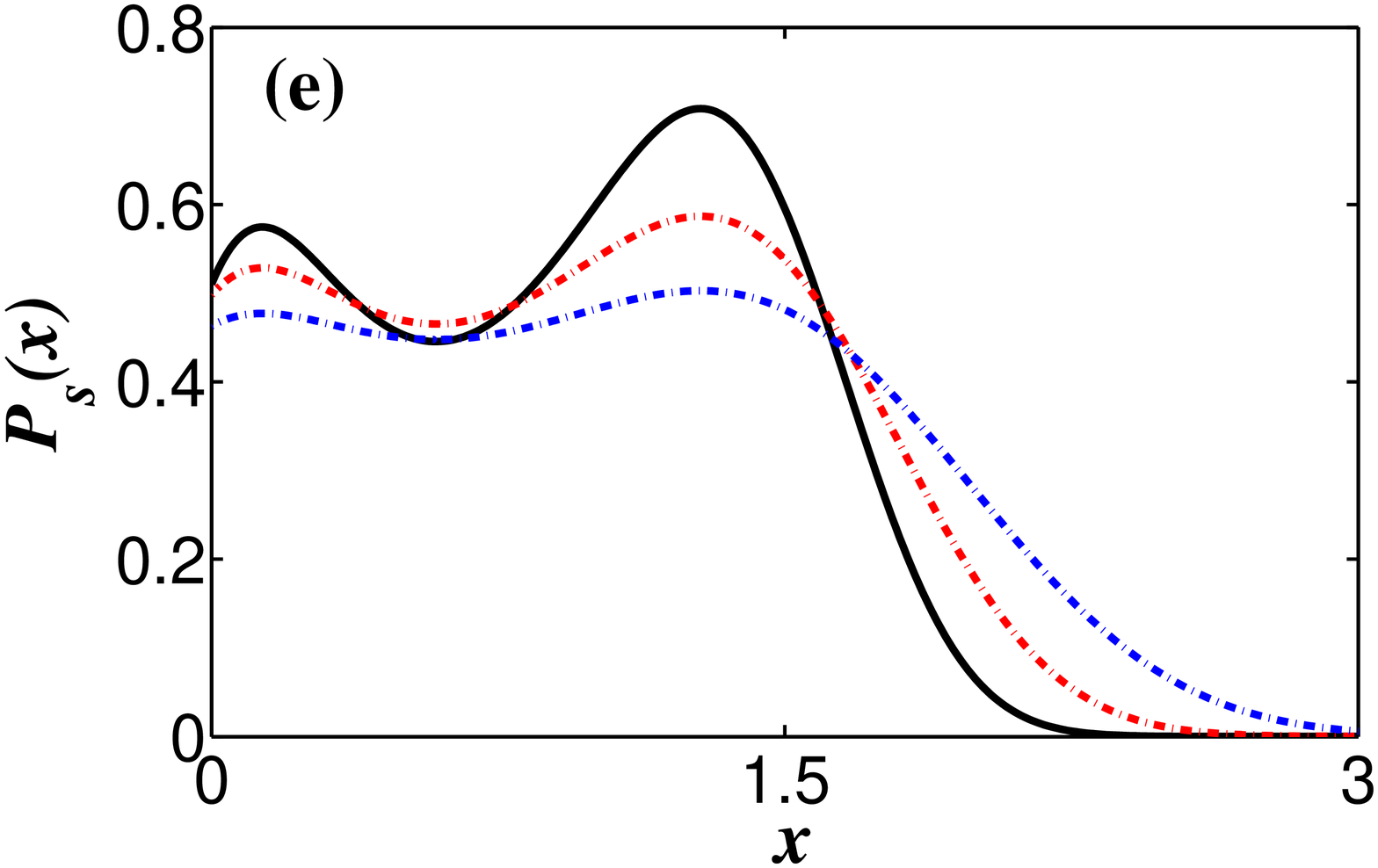}} \\
\resizebox{!}{1.08in}{\includegraphics{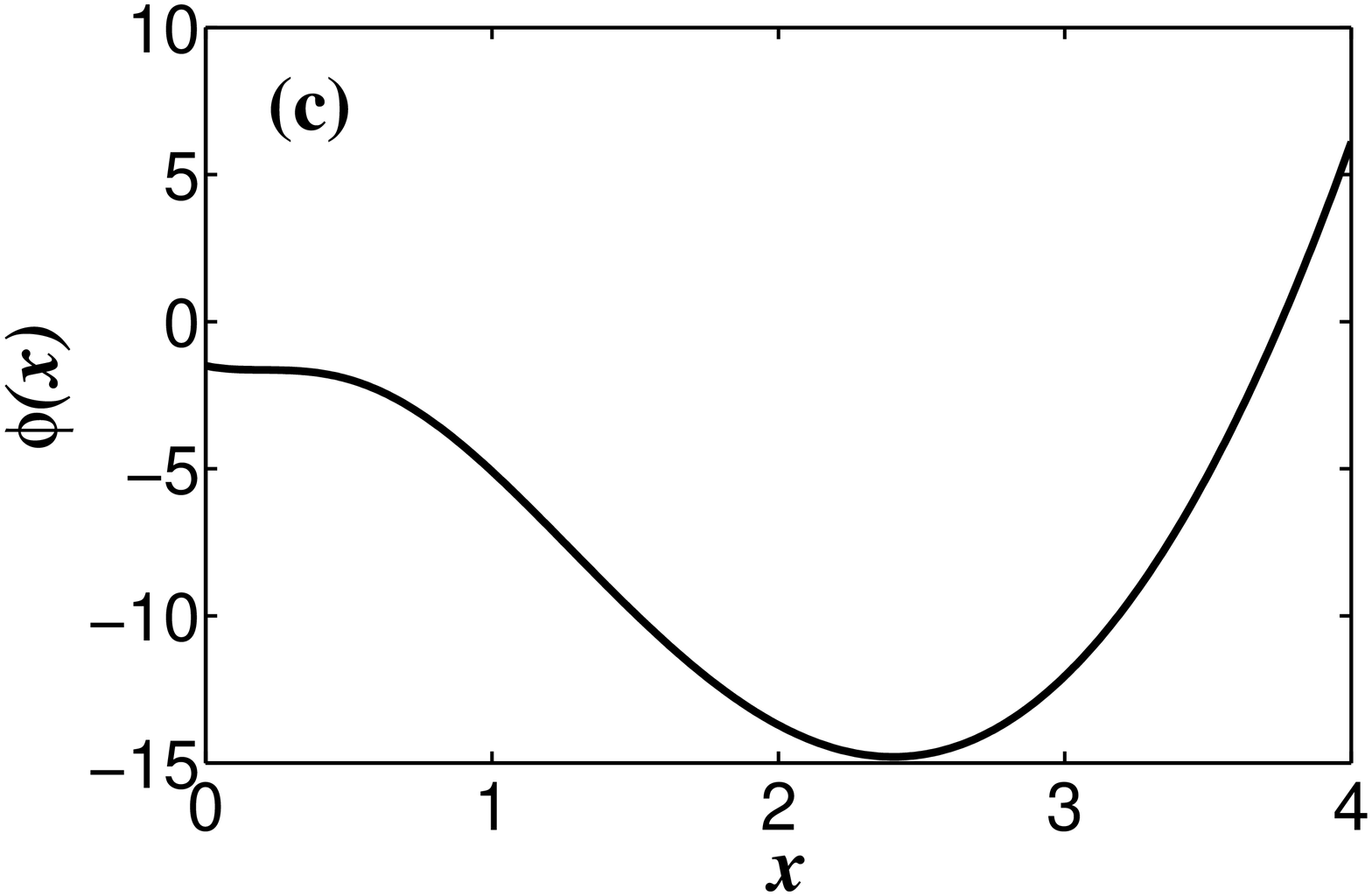}} & \hspace{-.2in}
\resizebox{!}{1.08in}{\includegraphics{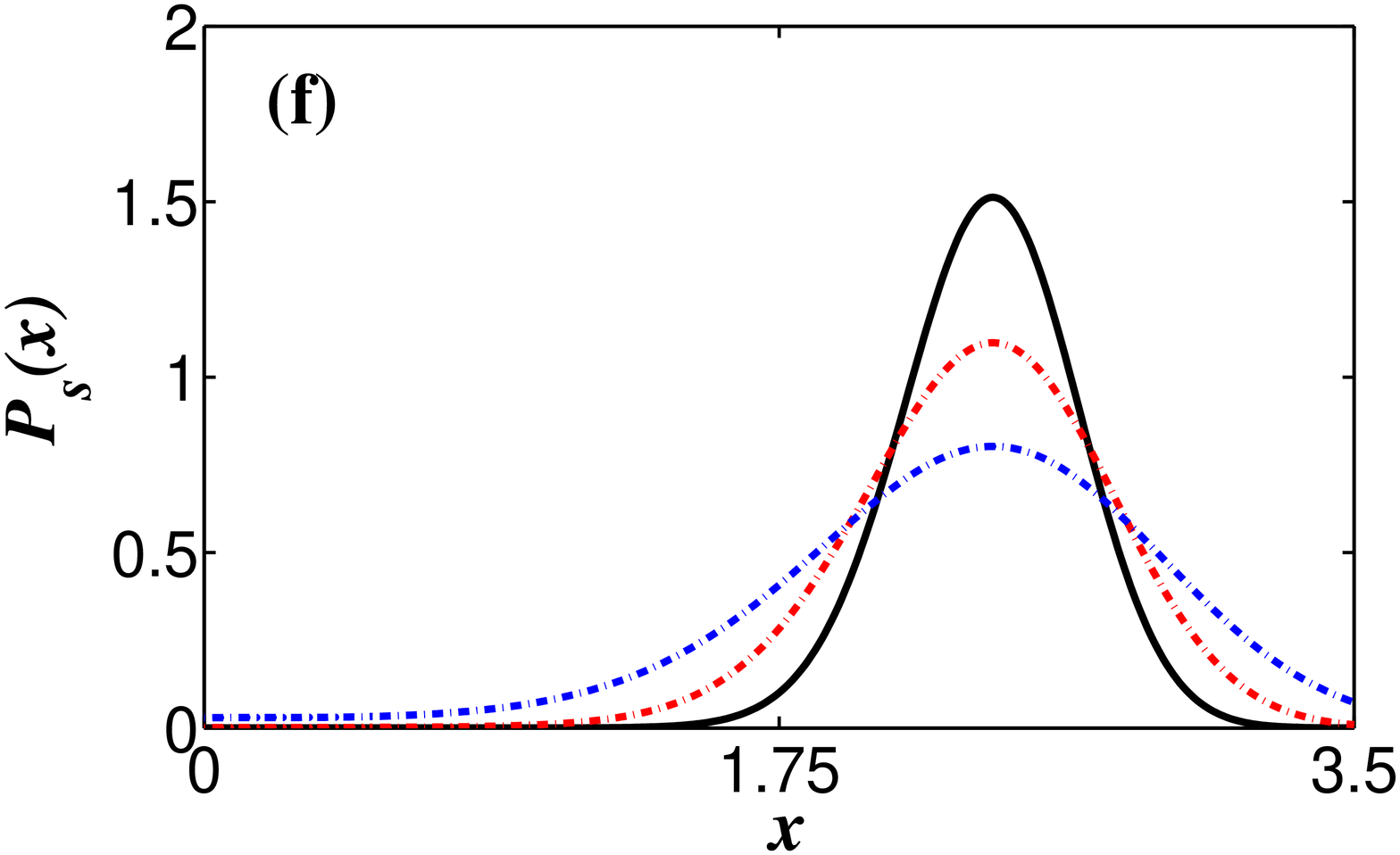}} 
\end{tabular}
\caption{\label{Fig:add} (a)--(c) The effect of maximum production
  rate $a$ on the stochastic potential $\phi(x)$, when only {\em
    additive noise} is present in the system: (a) $a=1.3$, (b)
  $a=1.9$, and (c) $a=2.7$ with $r=0.1$ and
  $\sigma_{2}=0.05$. (d)--(f) The effect of additive noise strength
  $\sigma_{2}$ together with maximum production rate $a$ on the SPDF
  $P_s(x)$: (d) $a=1.3$, (e) $a=1.9$, and (f) $a=2.7$ with $r=0.1$ and
  for three different noise strengths $\sigma_{2}=0.05$ (black curve)
  , $\sigma_{2}=0.1$ (red curve) and $\sigma_{2}=0.2$ (blue curve).}
\end{center}
\end{figure}

Figures~\ref{Fig:resili}(a) and \ref{Fig:resili}(d) show the effective
potential at two different values of $a$ with a non-zero additive
noise intensity.  The depth and width of a potential well can be
related to its {\em resilience}. Resilience is defined as the ability
of a system to recover to its original state upon a perturbation.  For
critical transitions as the system approaches to a bifurcation point
(see the positions of the parameter $a$ in Fig.~\ref{Fig:bif_PD}
marked by circles) the recovery rate from perturbation decreases
smoothly, known as critical slowing down and as a result the system
loses its resilience. One important prediction is that the loss of
resilience should lead to an increase in the standard deviation and
autocorrelation (see Figs.~\ref{Fig:resili}(c) and
\ref{Fig:resili}(f)) \cite{Scheffer:2012sc}. Next, we have
demonstrated how the probability distribution of a system can be
related with its potential and acts as a mapping between potential and
their resilience.  We simultaneously try to see the effect of changes
in the parameter $a$ and noise intensity on regime shifts.


The effect of changing the maximum production rate $a$ on the
stochastic potential $\phi(x)$ and simultaneously changing $a$
together with different additive noise intensity $\sigma_2$ on SPDF
$P_{s}(x)$ is depicted in Fig.~{\ref{Fig:add}}.  In the left panel of
Fig.~{\ref{Fig:add}}, we show that by varying $a$ from low (see
Fig.~\ref{Fig:add}(a)) to high values (see Fig.~\ref{Fig:add}(c)), the
system can pass from a monostable low protein concentration state
(Fig.~\ref{Fig:add}(a)) through a region of bistability
(Fig.~\ref{Fig:add}(b)) (i.e. coexistence of low and high
concentration states) to a monostable high concentration state
(Fig.~\ref{Fig:add}(c)).  For $a=1.9$, the region of bistability is
distinguished by the presence of two local minima in the potential
$\phi(x)$ (see Fig.~\ref{Fig:add}(b)).  As already stated, the depth
and width of the potential well determine the resilience of the
equilibrium point.  It is evident from Fig.~\ref{Fig:add}(b) that the
high concentration protein state ($x$) has high resilience.  Hence, in
a stochastic environment it has a low risk to experience a regime
shift and can sustain under large disturbance in comparison with the
low concentration protein state.  On the other hand in stochastic
models, peaks of the SPDF $P_{s}(x)$ correspond to attractors and
troughs correspond to repellors.  Moreover, an equilibrium point is
more stable (high resilience) if the SPDF peak is large in comparison
with another equilibrium point which is less stable (low resilience)
as the SPDF peak is small.  The effect of three different values of
noise intensity $\sigma_{2}$ on SPDF $P_{s}(x)$ is shown in
Figs.~\ref{Fig:add}(d)--\ref{Fig:add}(f) for three different values of
$a$.  In case of bistability, for a fixed $a=1.9$, if we slowly
increase the noise strength $\sigma_{2}$, the system still has coexistence
of low and high concentration protein state for low value of
$\sigma_{2}$.  As we further increase $\sigma_{2}$, the system will
slowly shift to a single state consists of high protein concentration
only (see Figs.~\ref{Fig:add}(e)--\ref{Fig:add}(f)).  This regime
shift is expected as because for $a=1.9$ the high protein
concentration state has high resilience.


\subsubsection{Multiplicative noise}

Next, we consider only the presence of multiplicative noise in the
system, i.e., in Eq.~(\ref{Eq:main}) the strength of additive noise
$\sigma_{2}=0$.  For $\sigma_{2}=0$, the FPE (\ref{Eq:FP1}) becomes:
\begin{equation}\label{Eq:FPMN}
\frac{\partial P(x,t)}{\partial t}=-\frac{\partial}{\partial x}\left[A(x)P(x,t)\right]
+\frac{\partial^2}{\partial x^2}\left[B(x)P(x,t)\right], 
\end{equation}
where
\begin{eqnarray*} 
A(x) & = & f(x)+\sigma_1g(x)g'(x),\;\; \text{and}\\ 
B(x) & = & \sigma_1\left[g(x)\right]^2 .
\end{eqnarray*}
The stationary solution $P_s(x)$ of Eq.~(\ref{Eq:FPMN}) is: 
\begin{equation}
P_{s}(x)=N_Me^{-\phi_(x)},
\end{equation}
where
\begin{equation} \label{eq:27}
\phi_(x)=\frac{1}{2}\ln\left[\sigma_{1}\left[g(x)\right]^{2}\right]
-\int^{x}\frac{f(x')dx'}{\sigma_{1}\left[g(x')\right]^2}\;,
\end{equation}
is the stochastic potential and $N_M$ is the normalization constant.
\begin{figure}[h!]
\begin{center}
\begin{tabular}{ll}
\resizebox{!}{1.08in}{\includegraphics{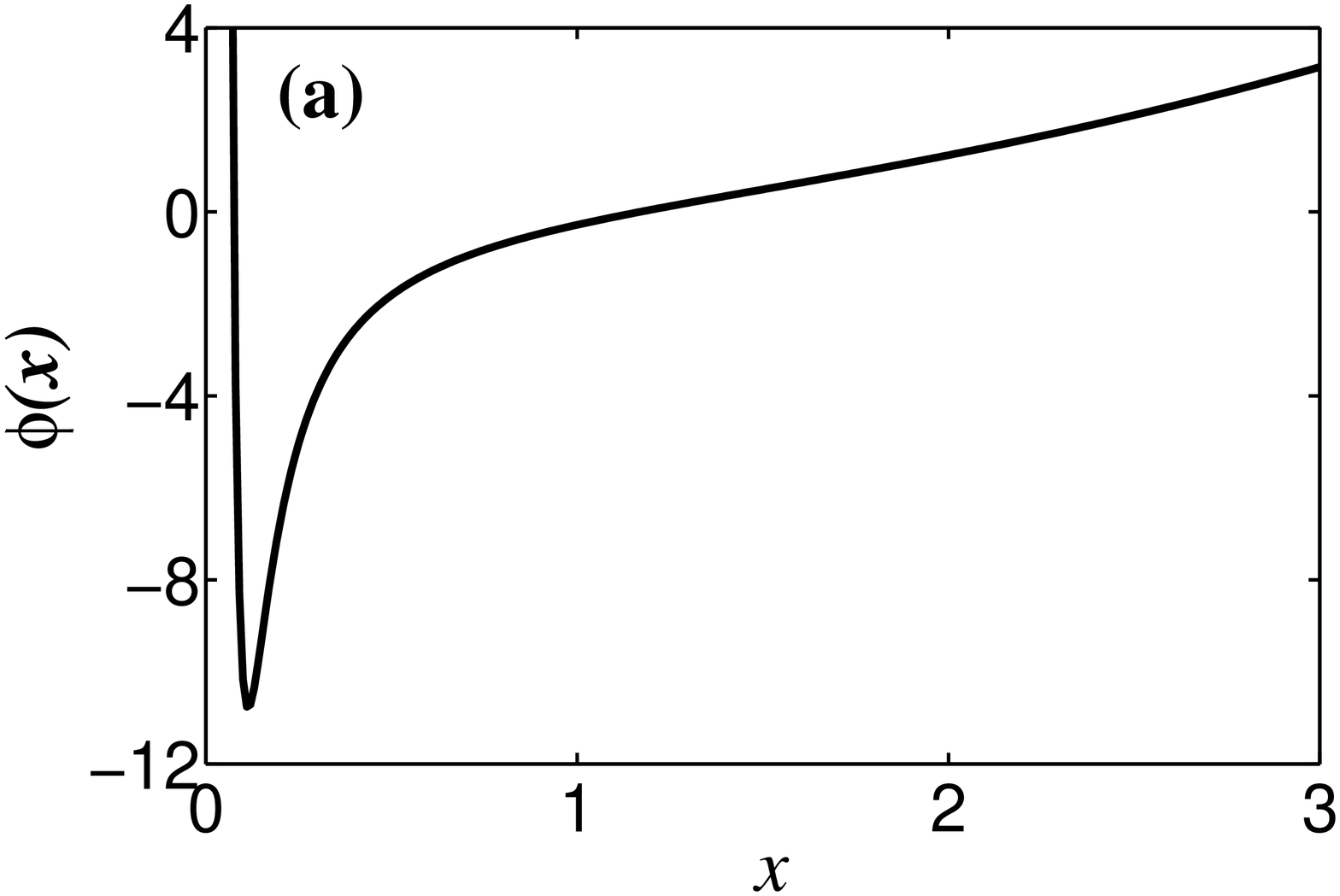}} \hspace{-.1in}  
\resizebox{!}{1.08in}{\includegraphics{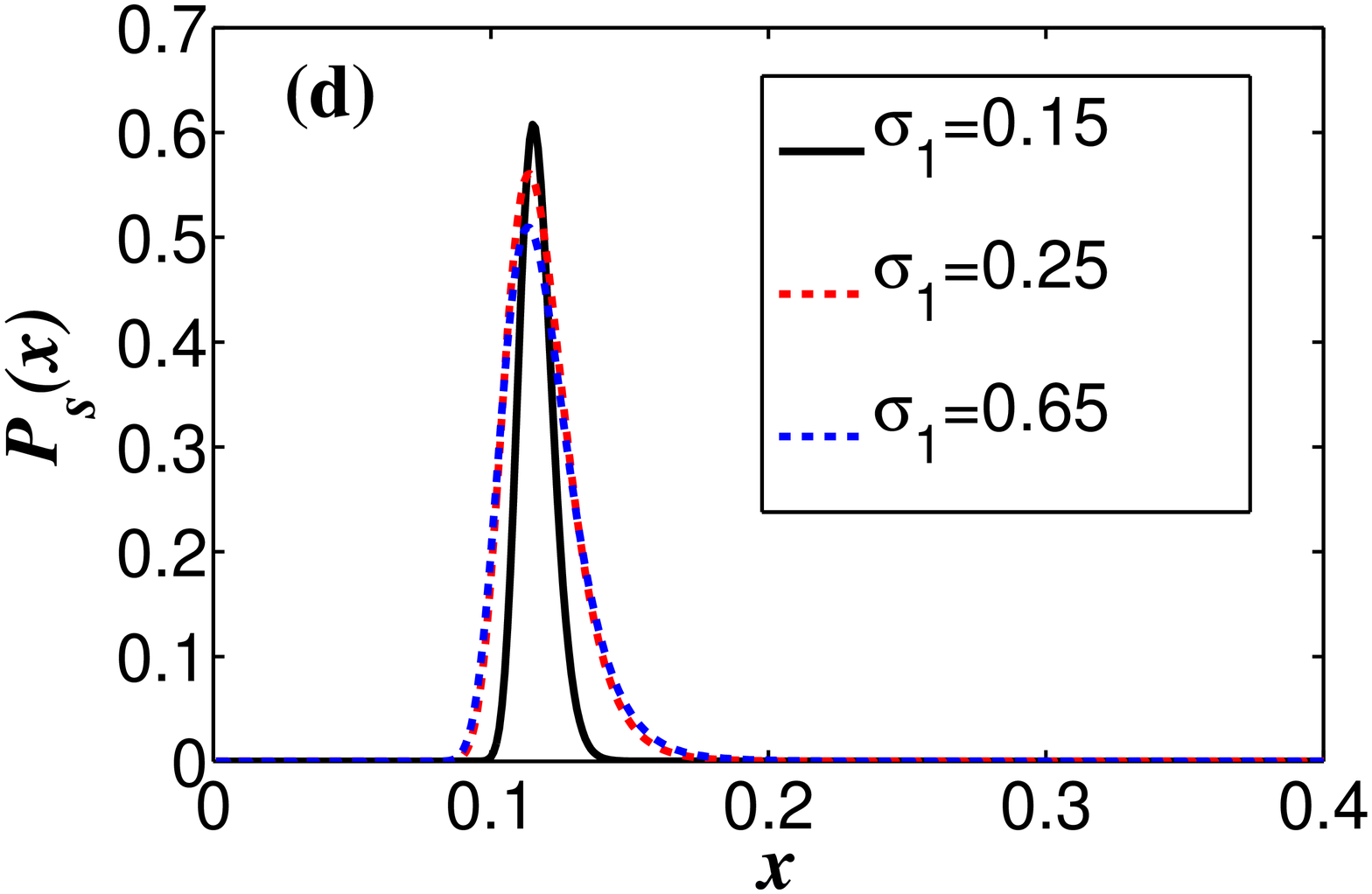}}\\
\resizebox{!}{1.08in}{\includegraphics{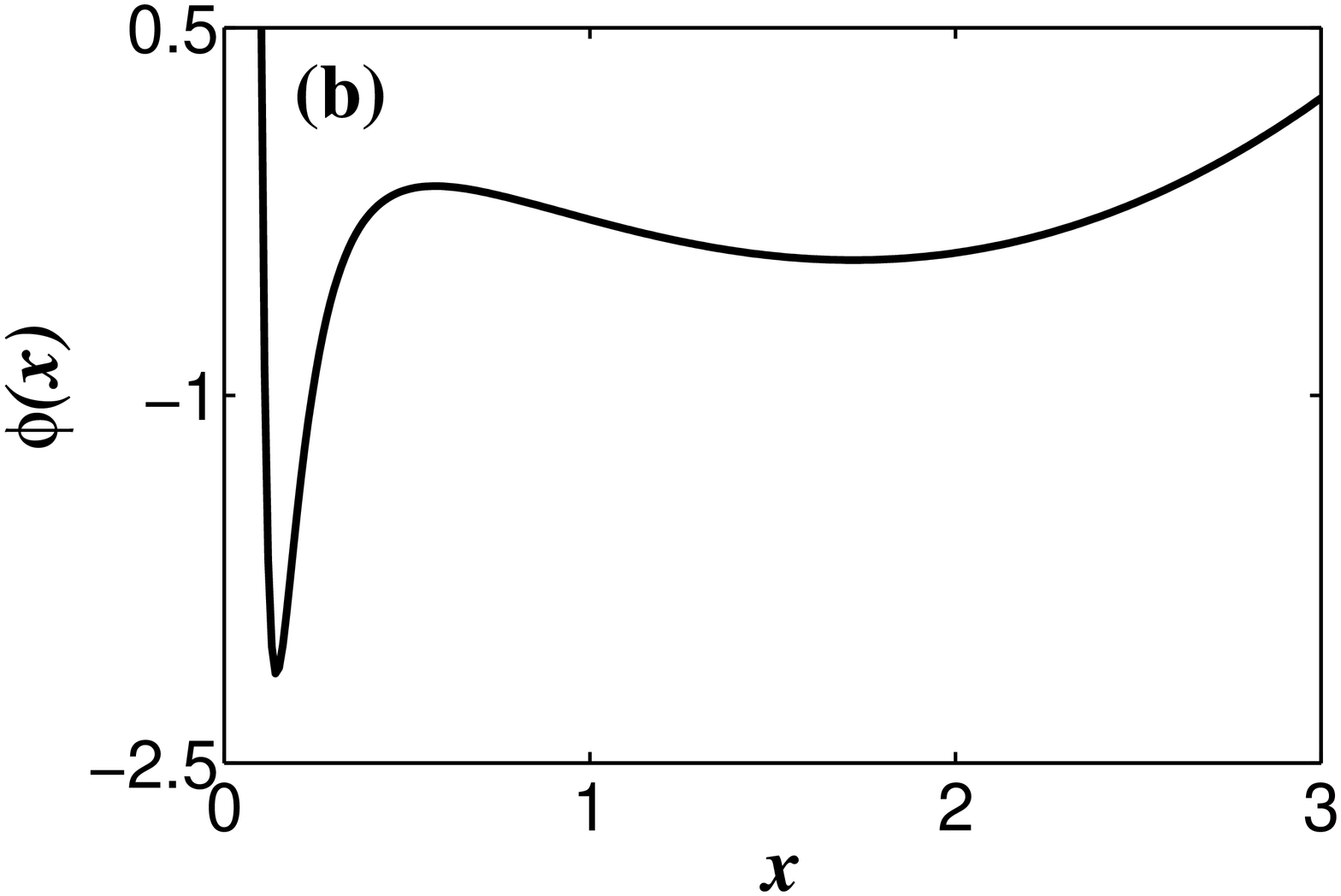}} \hspace{-.08in} 
\resizebox{!}{1.08in}{\includegraphics{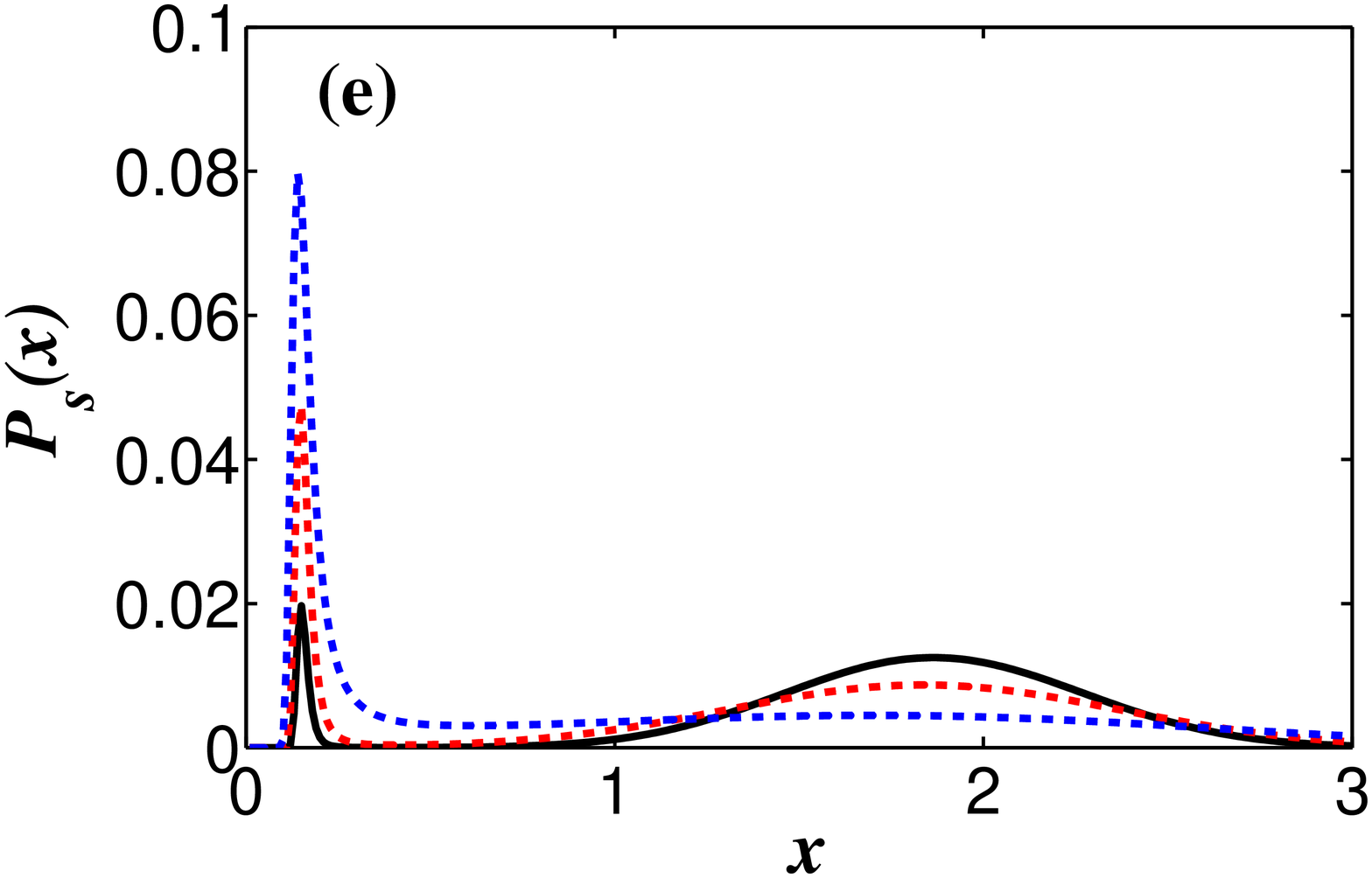}} \\
\resizebox{!}{1.1in}{\includegraphics{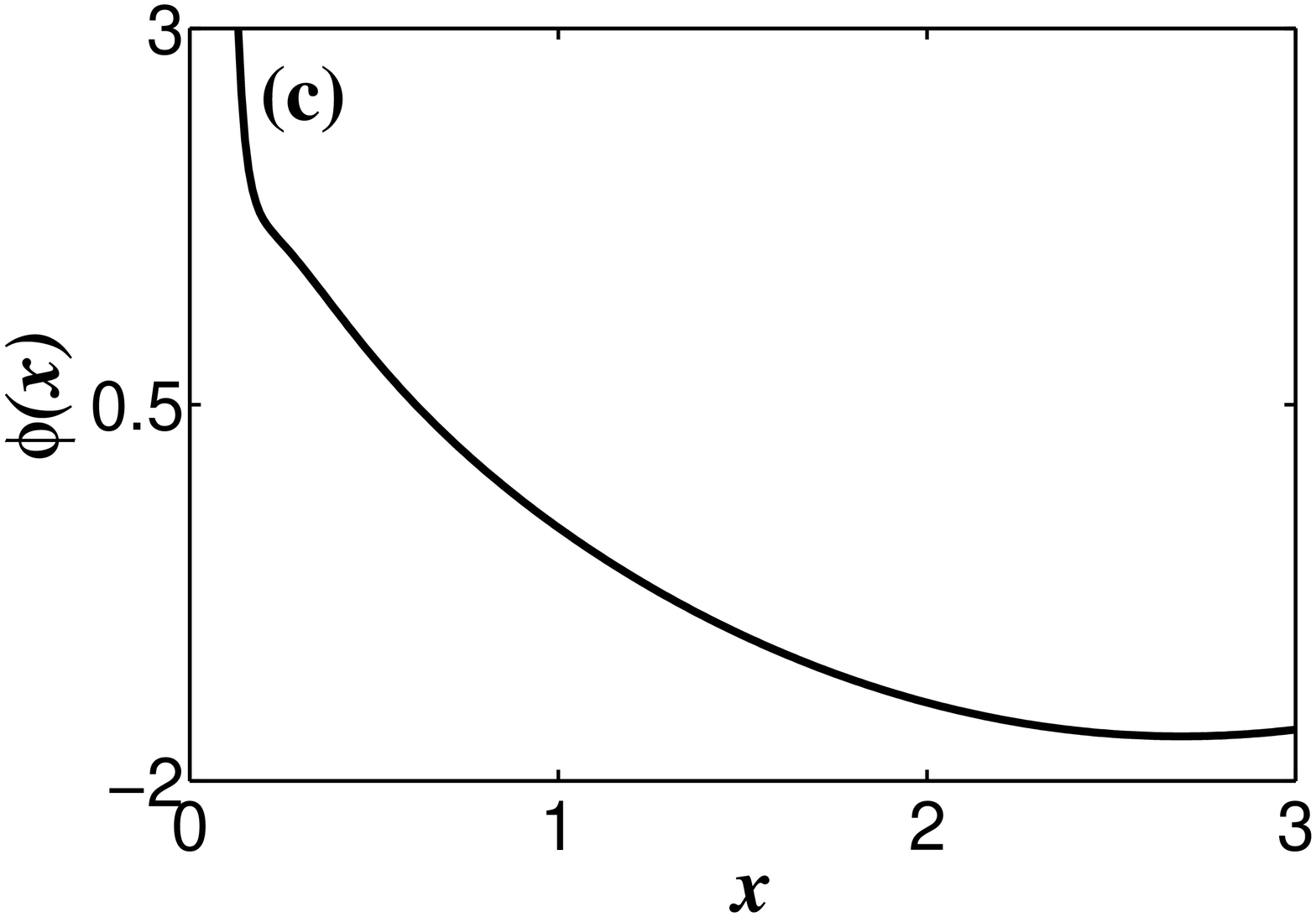}} \hspace{-.16in} 
\resizebox{!}{1.1in}{\includegraphics{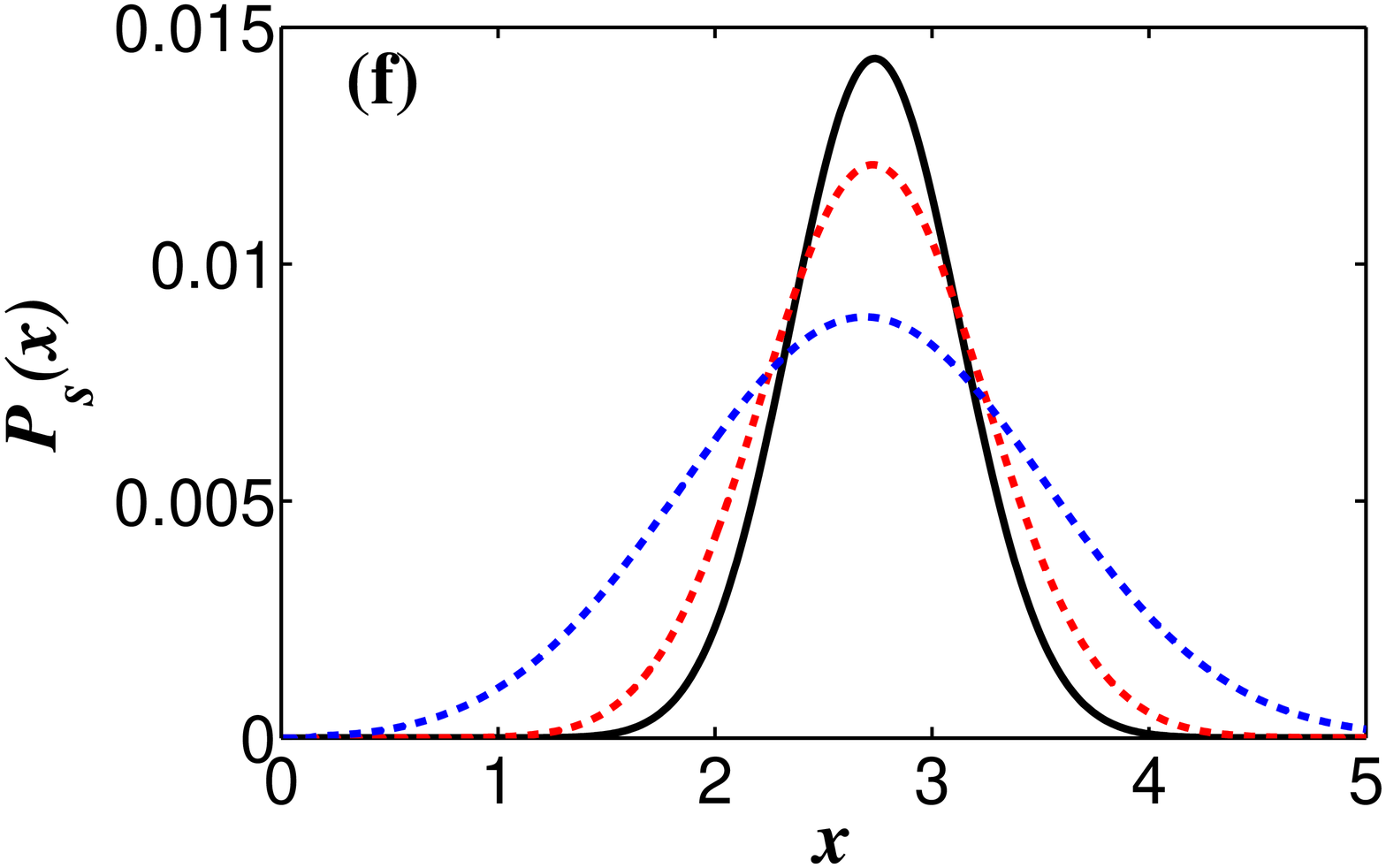}}
\end{tabular}
\caption{\label{Fig:mul} (a)--(c) The effect of maximum production
  rate $a$ on the stochastic potential $\phi(x)$, when only {\em
    multiplicative noise} is present in the system: (a) $a=1.2$, (b)
  $a=2.3$, and (c) $a=3$ with $r=0.1$, $\sigma_{1}=0.65$ and
  $\sigma_{2}=0$. (d)--(f) The effect of noise strength $\sigma_{1}$
  together with maximum production rate $a$ on the SPDF $P_s(x)$: (d)
  $a=1.2$, (e) $a=2.3$, and (f) $a=3$ with $r=0.1$ and for three
  different noise strengths $\sigma_{1}=0.15$ (black curve),
  $\sigma_{1}=0.25$ (red curve) and $\sigma_{1}=0.65$ (blue curve).}
\end{center}
\end{figure}

Similar to the case of additive noise in the previous section
(Sec.~\ref{SS:ANoi}), here Fig.~\ref{Fig:mul} shows the effect of
changing the parameter $a$ and the noise intensity $\sigma_{1}$ on the
potential function $\phi(x)$ and the SPDF $P_s(x)$. In one hand, when
we increase the parameter $a$ with a fixed noise intensity (see
Figs.~\ref{Fig:mul}(a)--\ref{Fig:mul}(c)), the system passes through a
low concentration state to a high concentration state via a bistable
region (Fig.~\ref{Fig:mul}(b)).  On the other hand, with variations in
$a$ we plotted the SPDF for different values of the noise intensity
$\sigma_{1}$ (see Figs.~\ref{Fig:mul}(d)--\ref{Fig:mul}(f)).  For
monostable regions the effect of noise intensity on $P_s(x)$ is
similar, with an increase in $\sigma_1$ the SPDF peak is decreasing.
However, in the bistable region (Fig.~\ref{Fig:mul}(e)), with an
increase in $\sigma_1$, the SPDF peak at the low protein concentration
$x$ is increasing and that of the high protein concentration $x$ is
decreasing.  This actually supports the fact that when there is
bistability in the system, the low concentration state is more
resilient (more stable under perturbative condition) than that of the
high concentration state, which is also evident from the
Fig.~\ref{Fig:mul}(d).


\subsubsection{Additive and multiplicative noise with correlation}

Herein, we consider that both the additive and multiplicative noise
$\eta(t)$ and $\xi(t)$ are present in the system Eq.~(\ref{Eq:main}).
Moreover, both these noise are statistically correlated with a cross
correlation parameter $\lambda$.  The correlation can arise from the
regulation of feedback mechanism, i.e., in the presence of noise the
protein concentration $x$ is chemically coupled to the transcription
rate $a$.  Here we discuss three cases corresponding to the degree of
correlation $\lambda$, which is considered negative, zero or positive.
In this section, we are mainly interested in to investigate how the
correlation between $\xi (t)$ and $\eta (t)$ triggers regime shifts
from high to low protein concentration state and vice versa. We also 
showed that by changing cross correlation parameter $\lambda$ one can 
regulate the production of protein levels .
\begin{figure}
\begin{center}
\begin{tabular}{ll}\hspace{-.05in}
\resizebox{!}{1.08in}{\includegraphics{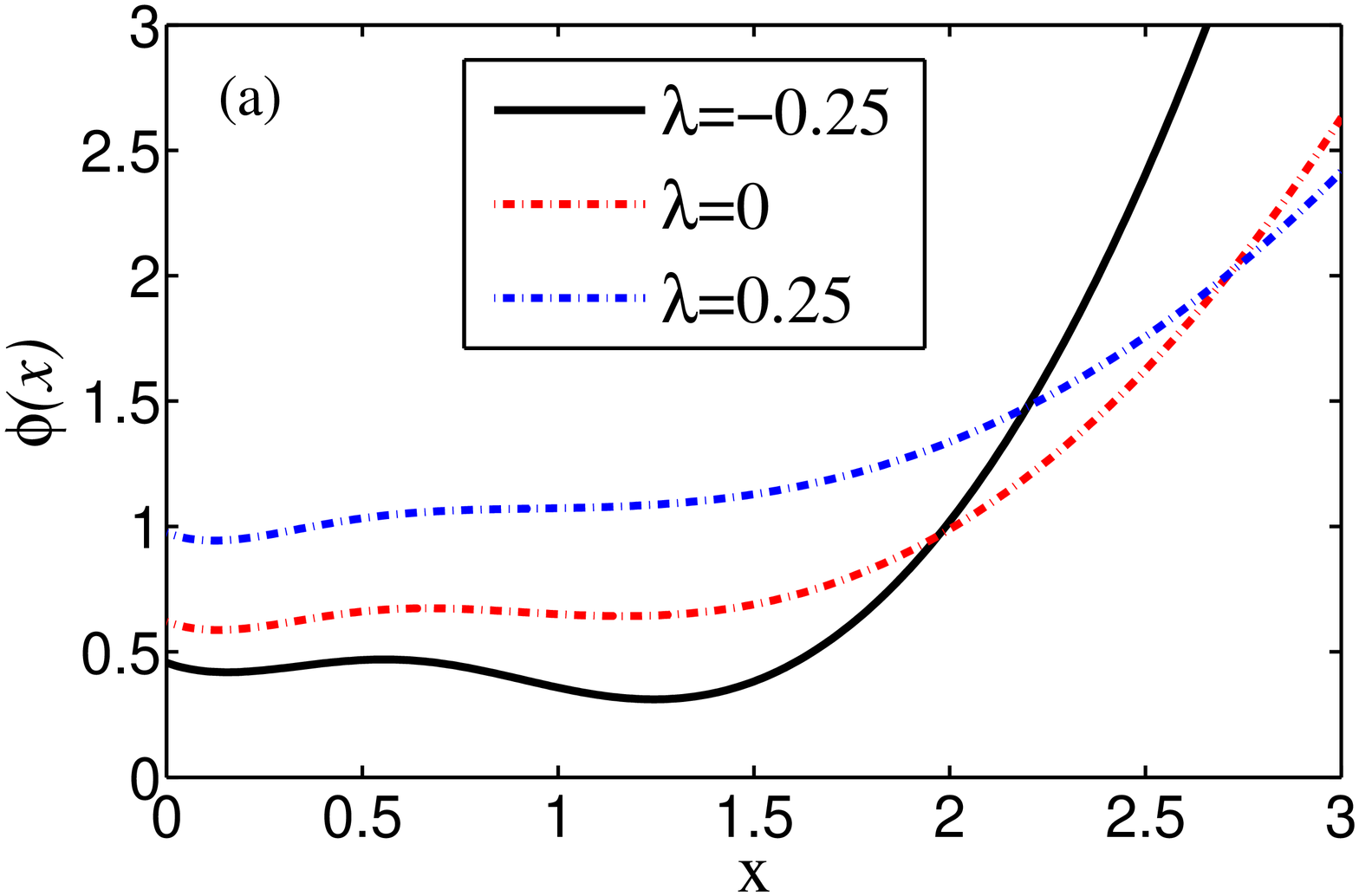}}  \hspace{-.22in}
\resizebox{!}{1.08in}{\includegraphics{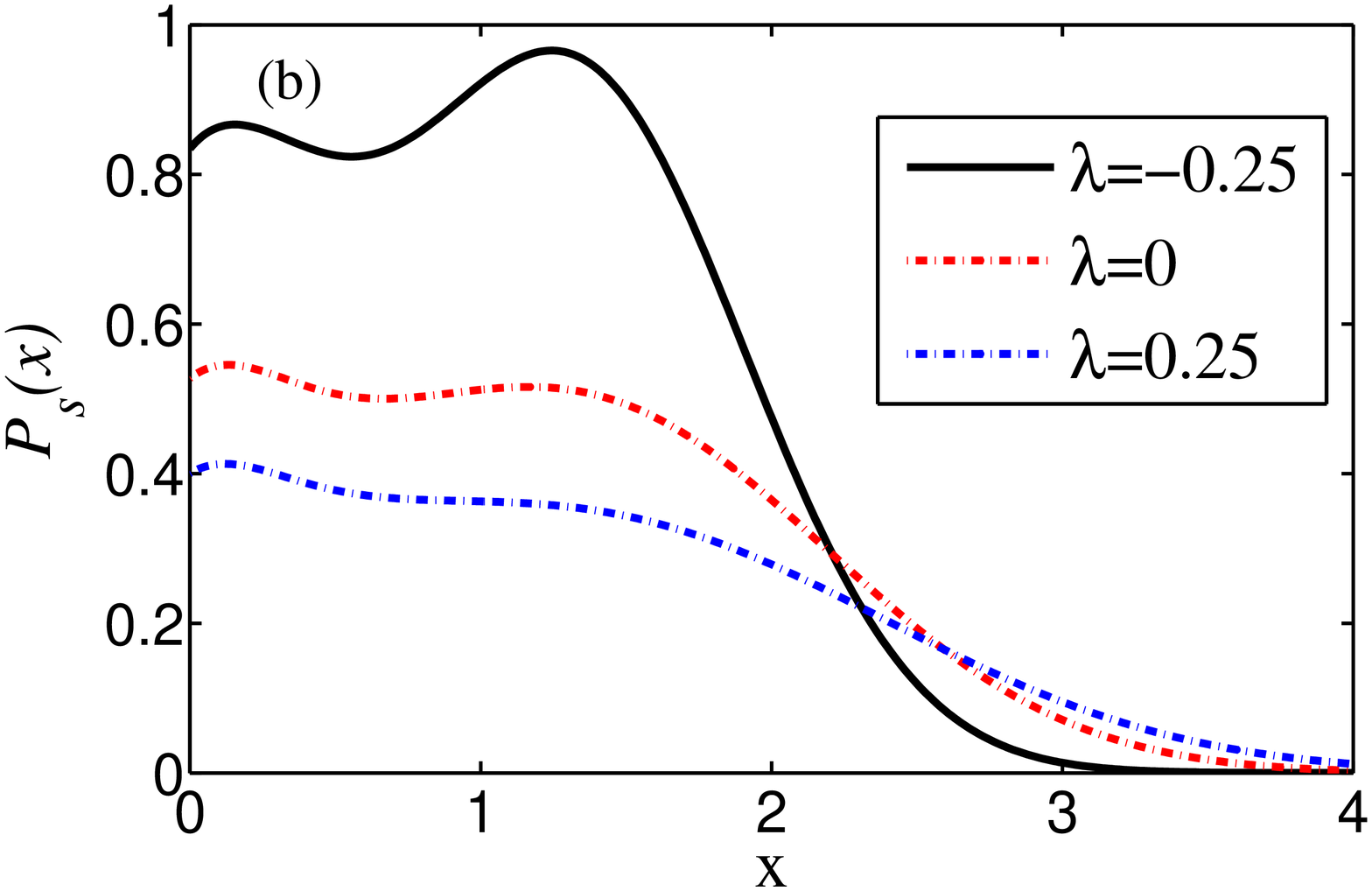}}  \\
\end{tabular}
\caption{\label{Fig:corr} (a) The potential $\phi_(x)$ and (b) the
  SPDF $P_{s}(x)$ for three different values of $\lambda$ : $-0.25$,
  $0$ and $+0.25$. The other parameter values are $r=0.1$,
  $a=1.9$, $\sigma_1=0.4$ and $\sigma_2=0.9$.}
\end{center}
\end{figure}

The expressions of the SPDF $P_s(x)$ and the potential $\phi(x)$ are
given in Eq.~(\ref{Eq:SPDF1}) and Eq.~(\ref{Eq:POT1}), respectively.
Figure~\ref{Fig:corr} shows the results for $\phi(x)$ and $P_s(x)$ for
three different strengths of correlation between additive and
multiplicative noise.  The system parameters $r$ and $a$ are chosen in
such a way that the deterministic system retains bistability.  As we
vary the correlation parameter $\lambda$ from a negative value to a
positive value, it is evident from Fig.~\ref{Fig:corr}(a) that the
resilience of right potential well is more sensitive with the change
in $\lambda$ than that of left potential well.  In fact, negative
correlation ($\lambda < 0$) will increase the stability of the high
concentration of protein and positive correlation ($\lambda > 0$) will
increase the stability of the low concentration of protein (see the
SPDF in Fig.~\ref{Fig:corr}(b)).  Hence, an increase in the cross
correlation intensity always induces regime shifts from high to low
protein concentration state.


\subsection{Mean first-passage time}
\label{sec4}

In biological systems, where noise plays an important role in
determining the dynamics, it is of general interest to calculate the
robustness of the system steady state under perturbative condition.
The robustness of a steady state can be quantified by the mean
first-passage time (MFPT) in noise driven systems \cite{Ga09,Dr07}.
In the bistable potential of Eq.~(\ref{Eq:main}), let
$x^{st}_l/x^{st}_u$ $(x^{st}_l<x^{st}_u)$ be the two steady states
corresponding to the low/high protein concentrations separated by the
potential barrier $x^{un}_b$ (i.e., the unstable equilibrium point).
An equilibrium point can exit from its potential well in the presence
of noise.  The exit time is a number which depends on the specific
realization of the random process and is known as first passage time.
When the first passage time is averaged over many realizations, we get
the mean first passage time.

\begin{figure}[h!]
\begin{center}
\begin{tabular}{l}
\resizebox{!}{1.4in}{\includegraphics{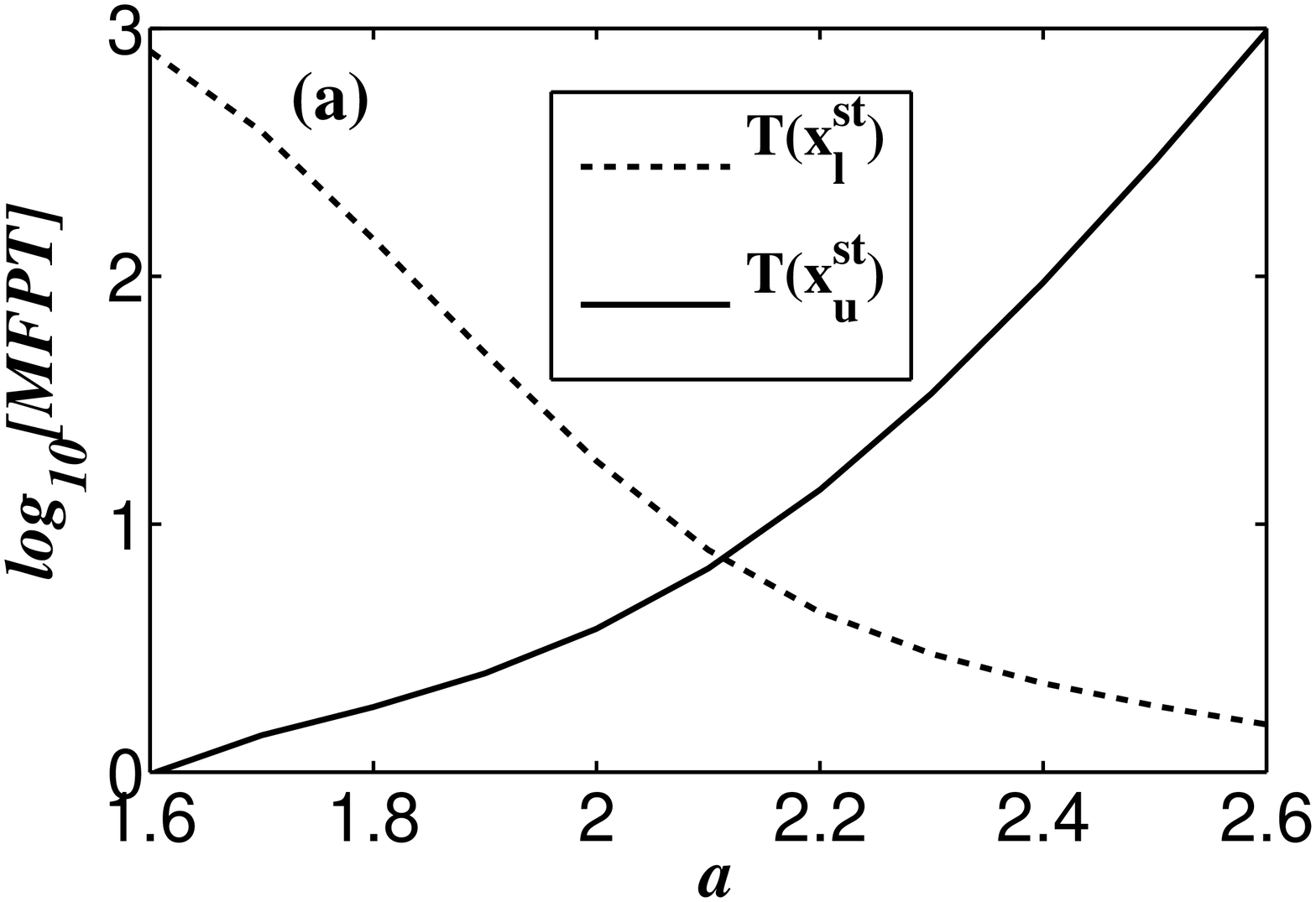}}\\
 \hspace{-0.09in}  \resizebox{!}{1.42in}{\includegraphics{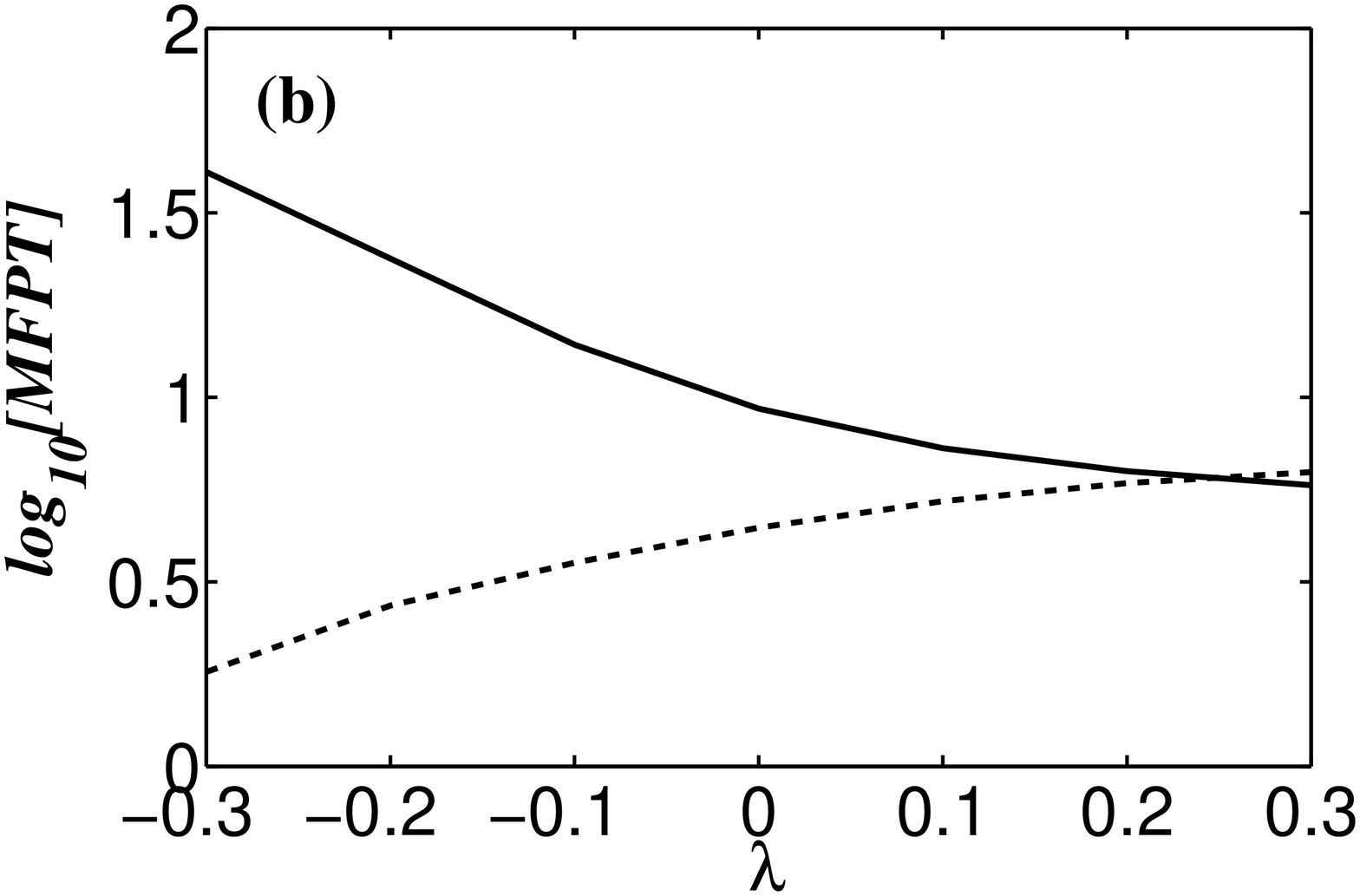}}
\end{tabular}
\caption{\label{MFPT} (a) The logarithmic value of MFPT with an
  increase in $a$.  The other parameter values are $r=0.1$,
  $\sigma_{1}=0.2$, $\sigma_{2}=0.9$ and $\lambda=0.01$.  (b) The
  logarithmic value of MFPT with an increase in $\lambda$. The parameter
  values are $r=0.1$, $a=1.9$, $\sigma_{1}=0.4$ and $\sigma_{2}=0.9$.}
\end{center}
\end{figure}
In the context of anticipating regime shifts, MFPT provides a very
useful characterization of the time scale on which a critical
transition is likely to happen. When the structure of a dynamical
system is known, the early warning signal indicators of
regime shifts should be measured from the time series data of length
shorter than the MFPT of the state in that particular regime.

\begin{figure}[h!]
\begin{center}
\begin{tabular}{l}
\resizebox{!}{1.5in}{\includegraphics{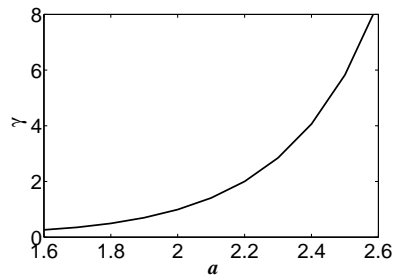}}
\end{tabular}
\caption{\label{MFPT1} Plot of the ratio $\gamma = \frac{T(
    x^{st}_u)}{T( x^{st}_l)}$ of the MFPT with respect to the control
  parameter $a$.  The other parameter values are $r=0.1$,
  $\sigma_{1}=0.4$ and $\sigma_{2}=0.9$ and $\lambda=0.1$.}
\end{center}
\end{figure}
\begin{figure*}[!ht]
\begin{center}
\begin{tabular}{lll}
{\footnotesize (a)} & {\footnotesize (b)} & {\footnotesize (c)}
\\ \resizebox{!}{1.54in}{\includegraphics{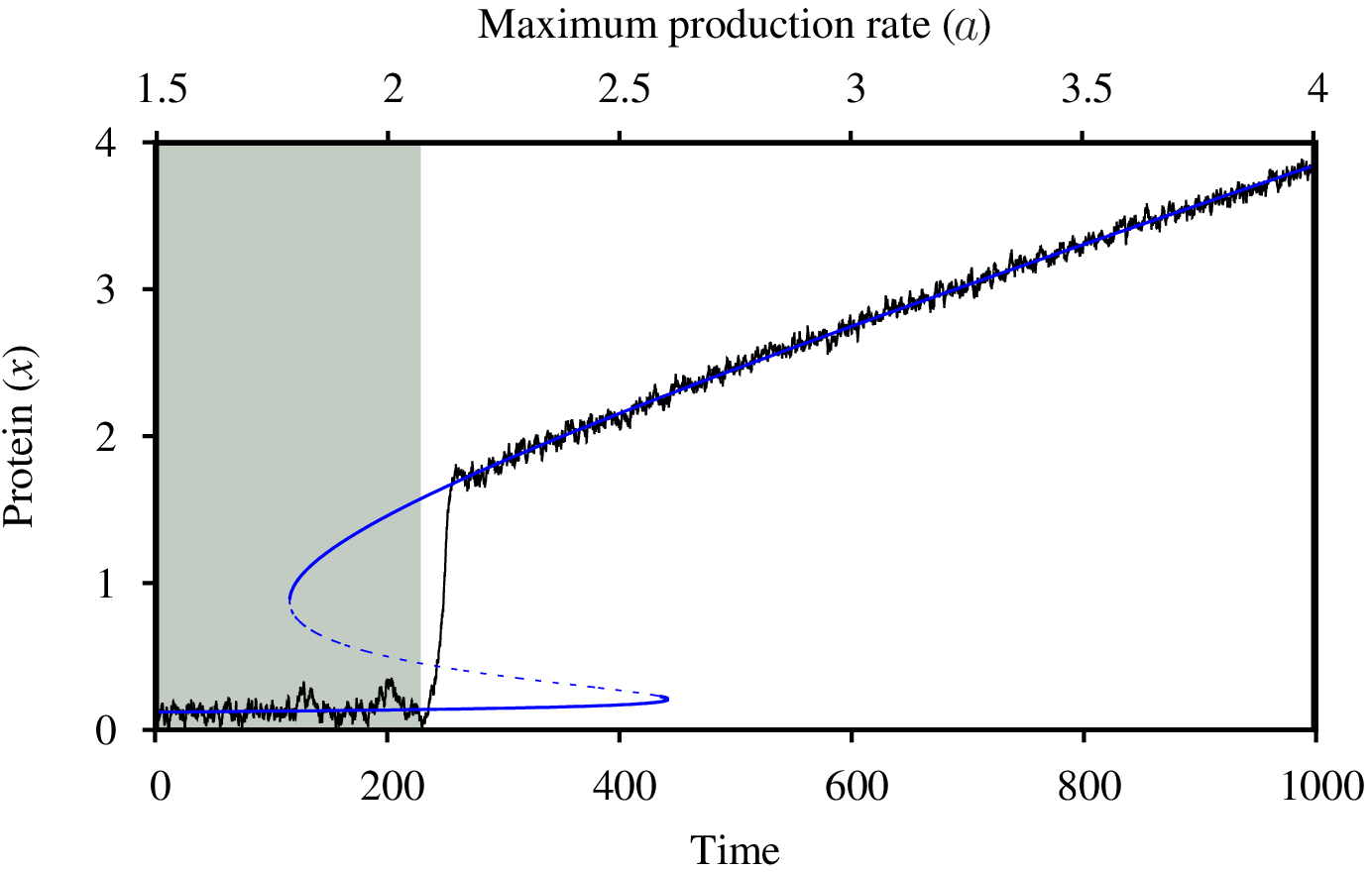}}\hspace{-0.09in}
& \resizebox{!}{1.54in}{\includegraphics{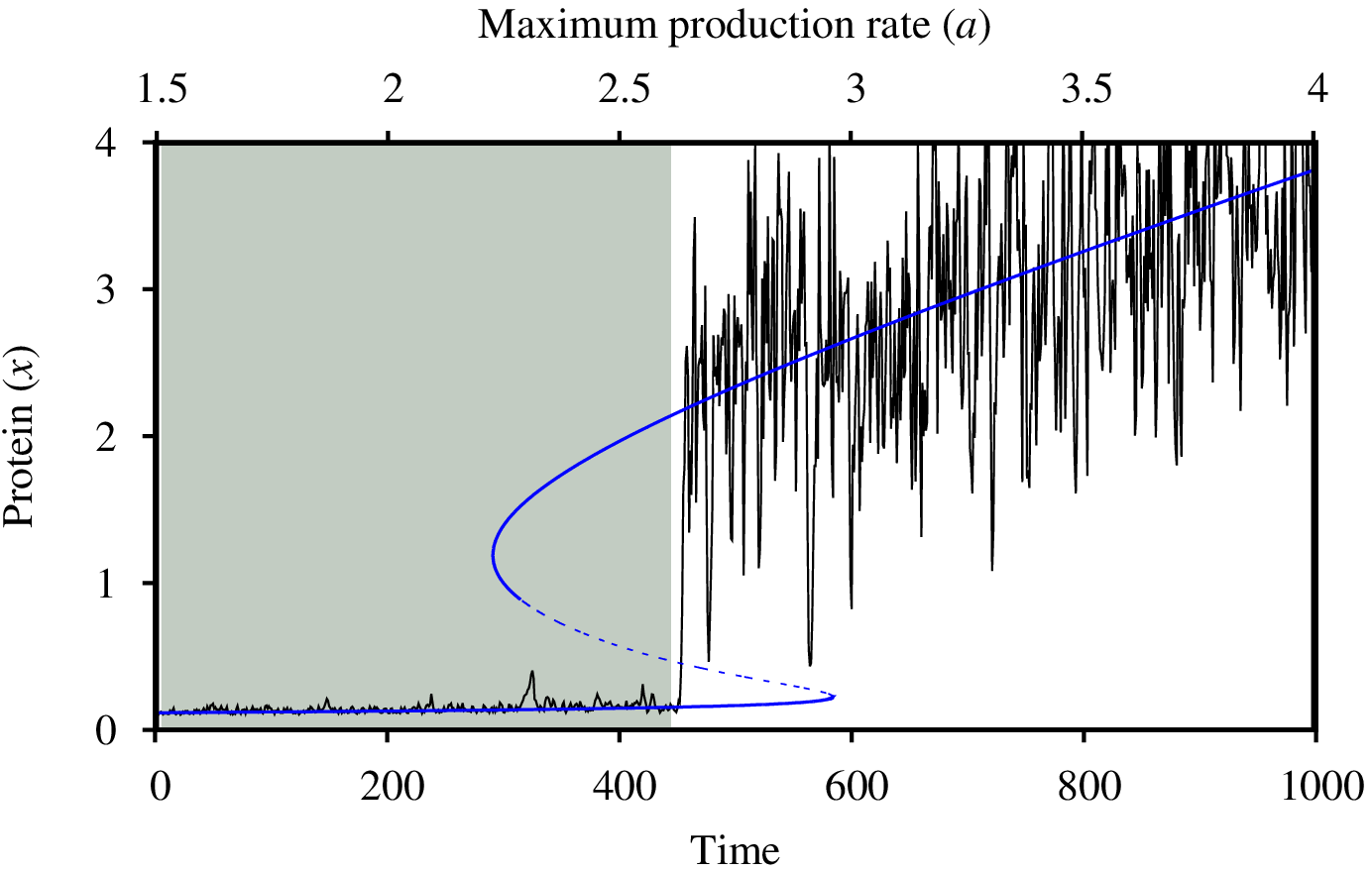}}
\hspace{-0.09in} &
\resizebox{!}{1.54in}{\includegraphics{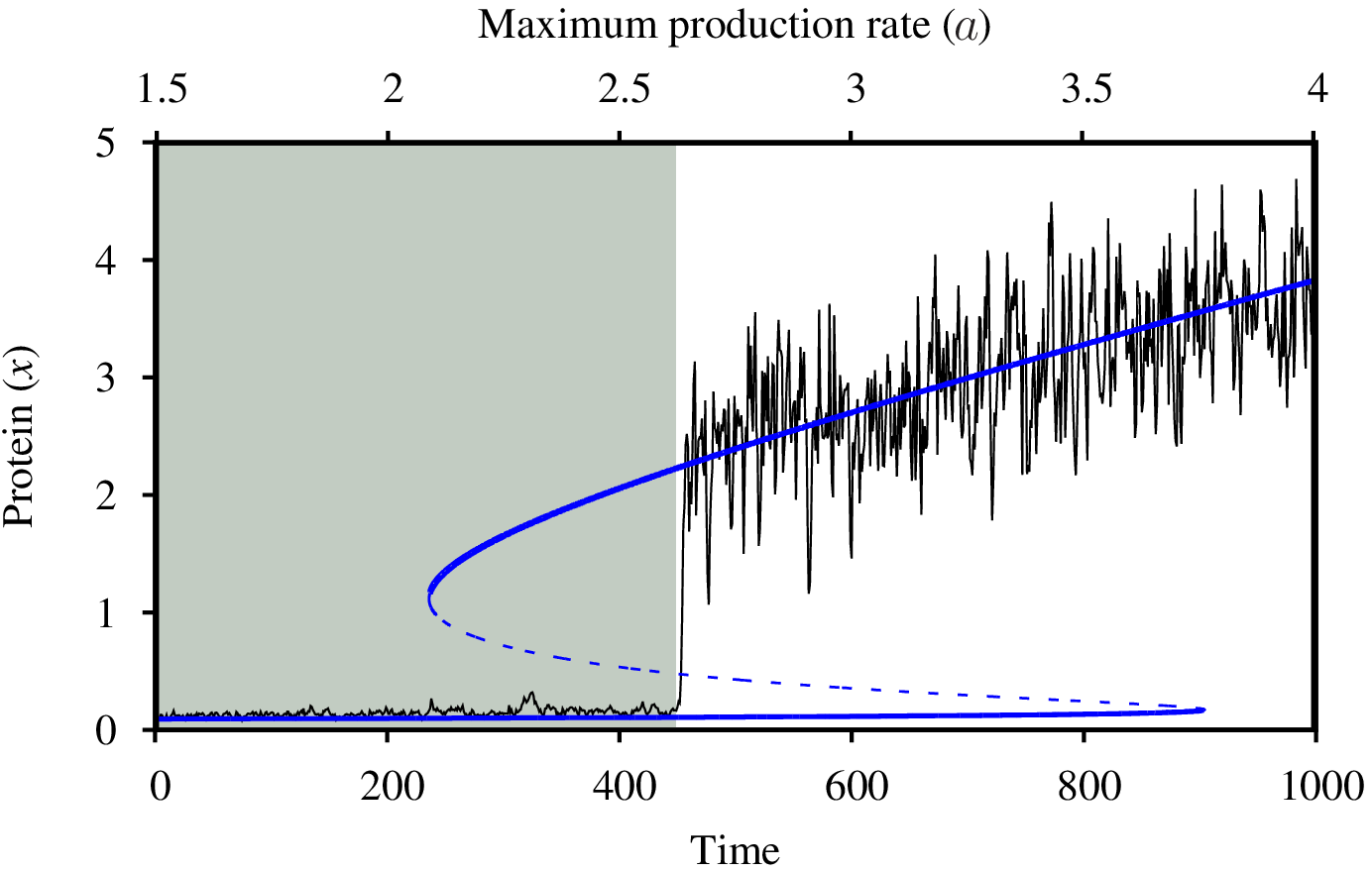}} \vspace{0.09in}
\\ \resizebox{!}{2.3in}{\includegraphics{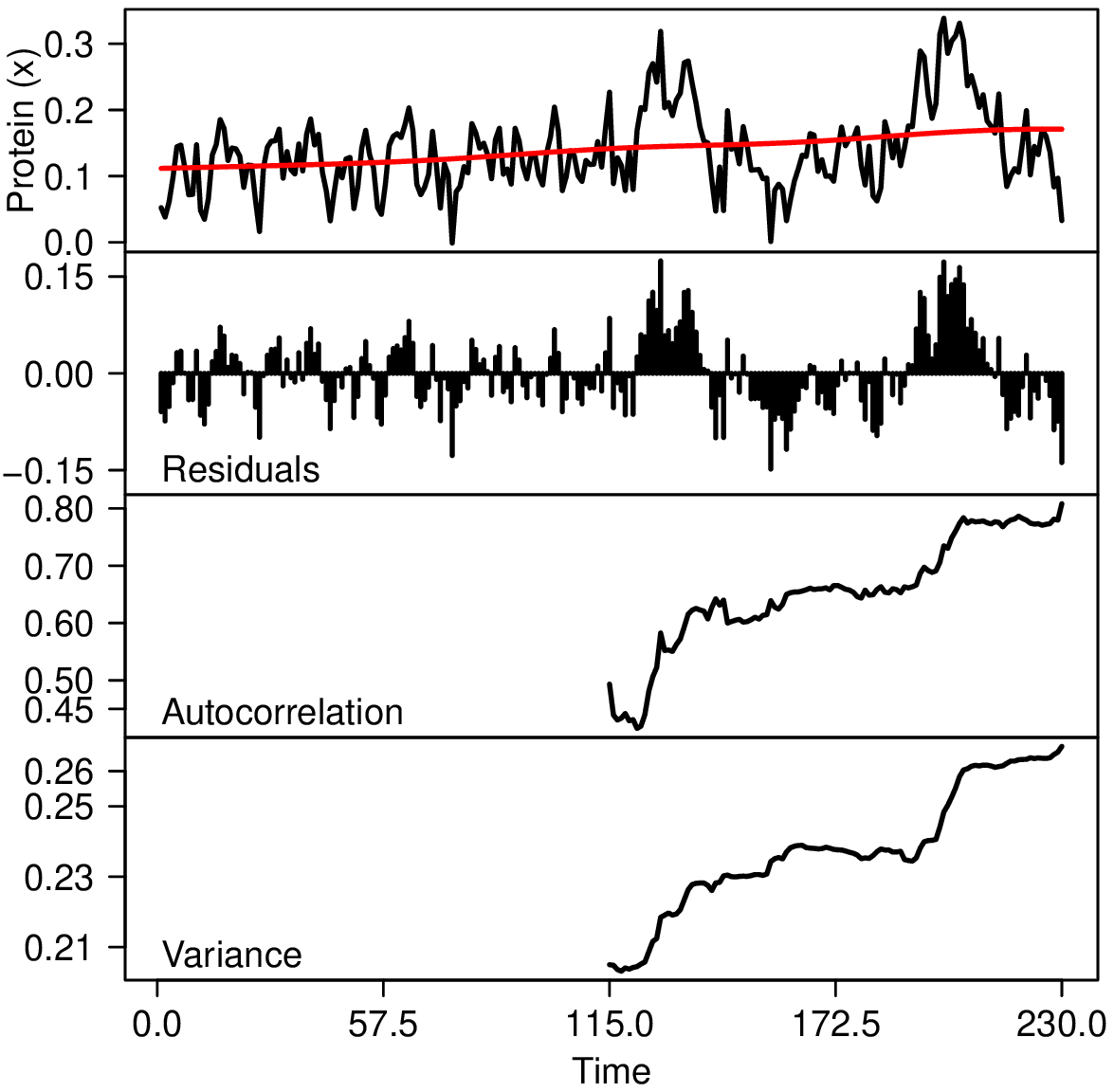}}\hspace{-1.2in}
& \resizebox{!}{2.3in}{\includegraphics{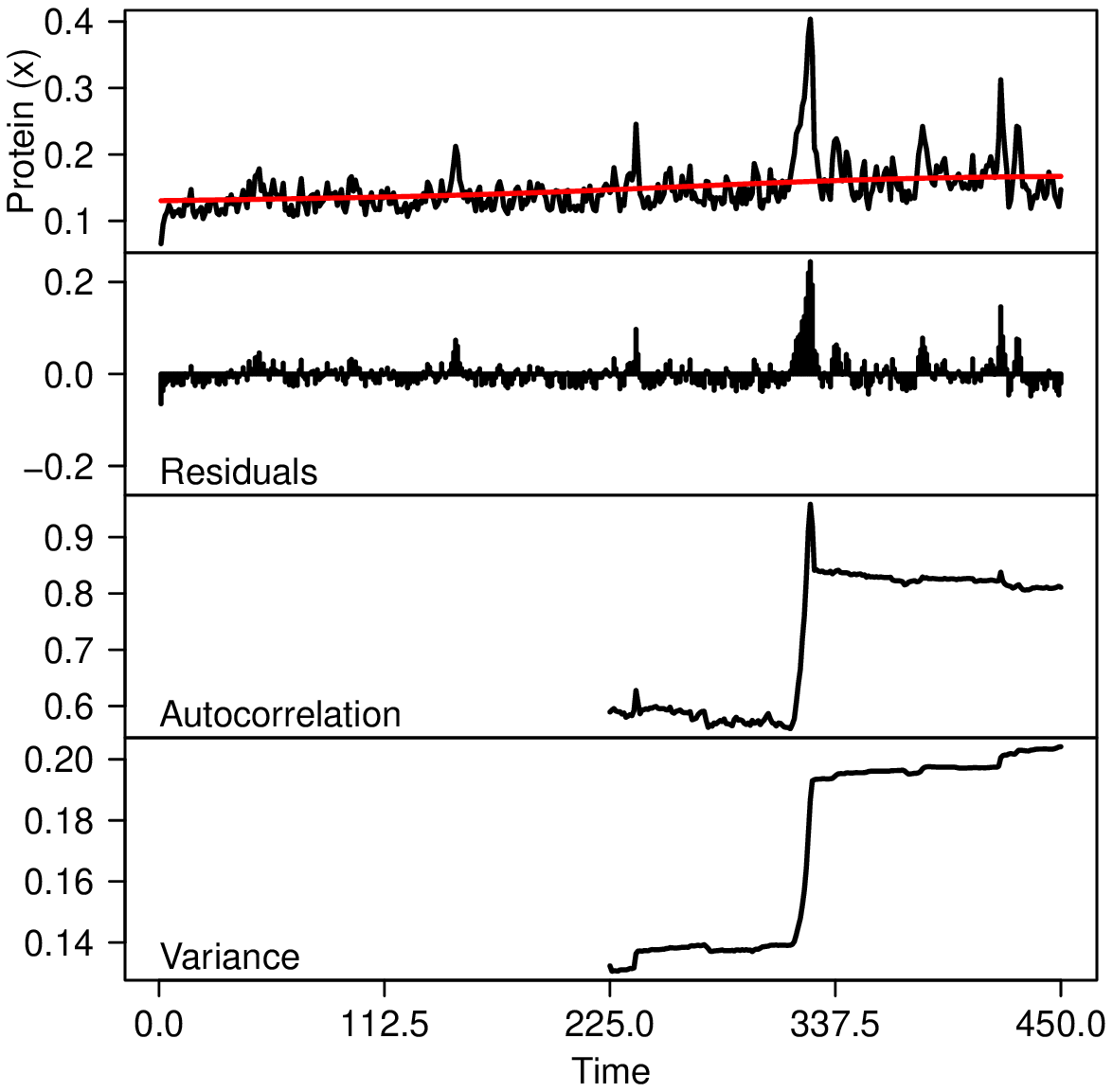}}
\hspace{-1.2in} & \resizebox{!}{2.3in}{\includegraphics{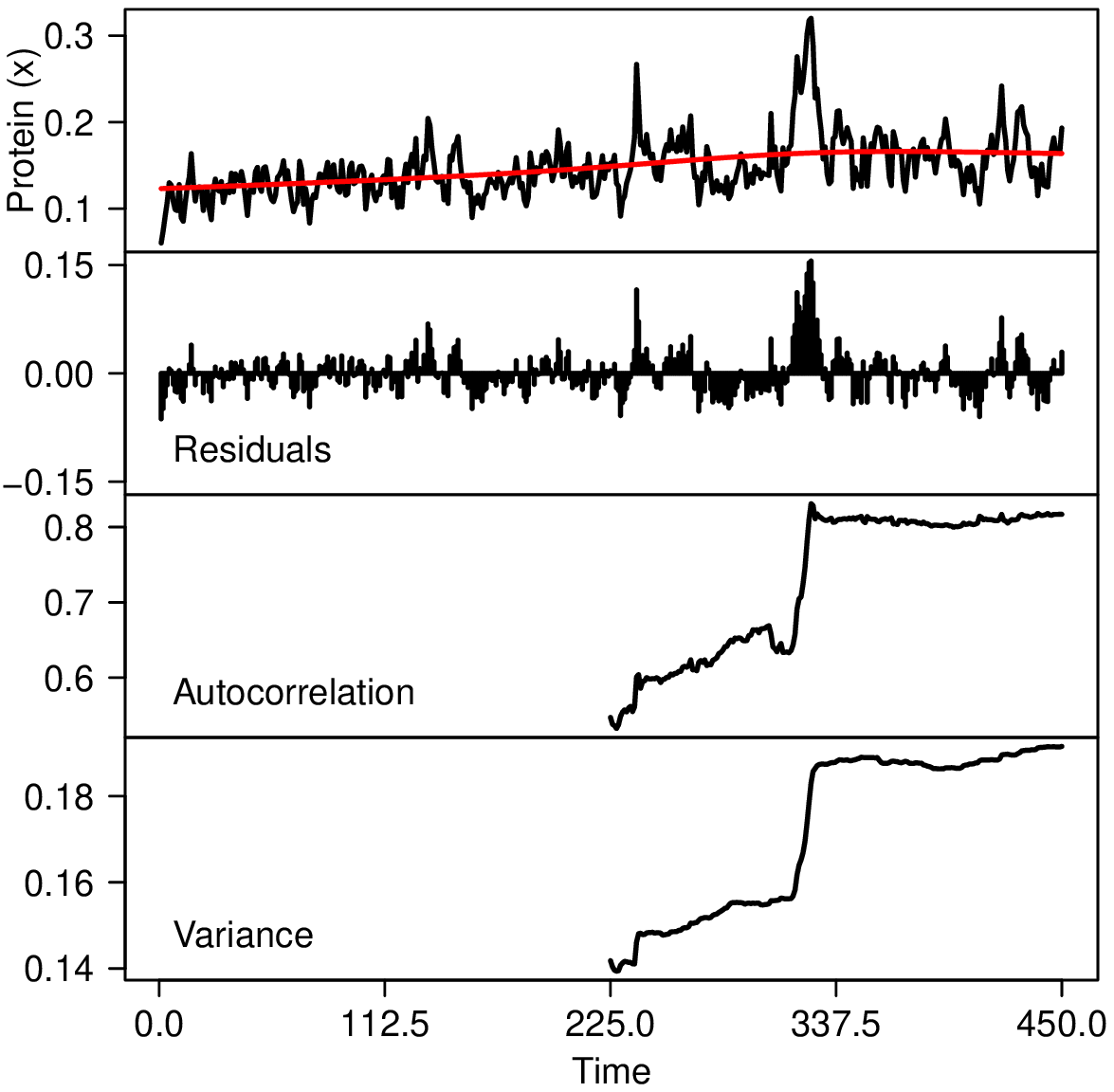}}  \\
\end{tabular}
\caption{\label{csd} Early warning signals for simulated time series
  data of the stochastic model in the case of critical slowing
  down. We calculated variance and autocorrelation within rolling
  window of half the length of the time series segment (a segment is
  indicated by shaded region): (a) Additive noise: parameter values
  are $r=0.1$, $\sigma_{1}=0$ and $\sigma_{2}=0.9$; (b) Multiplicative
  noise: parameter values are $r=0.1$, $\sigma_{1}=0.9$ and
  $\sigma_{2}=0$; and (c) Correlated noise: parameter values are
  $r=0.1$, $\sigma_{1}=0.4$, $\sigma_{2}=0.1$ and $\lambda=0.5$.  See
  text for more details.}
\end{center}
\end{figure*}

It is clear that the basin of attraction of the state $x^{st}_u$
extends from $x^{un}_b$ to $+\infty$, as it is in the right of
$x^{st}_l$.  Let $T(x)$ be the MFPT to state $x^{un}_b$ starting at $x
> x^{un}_b$. Then $T(x)$ satisfies the following ordinary differential
equation \cite{Ga09}:
\begin{equation}\label{Eq:mft}
A(x)\frac{\partial T(x)}{\partial x} + \frac{1}{2}B(x)\frac{\partial
  ^2T(x)}{\partial x^2}= -1,
\end{equation}
with boundary conditions $T(x^{un}_b)=0$ and
$\frac{\partial T(+\infty)}{\partial x}=0$.  Similarly one can calculate the MFPT to
state $x^{un}_b$ for the basin of attraction of the state $x^{st}_l$
which extends from $0$ to $x^{un}_b$.  After solving
Eq.~(\ref{Eq:mft}), one can obtain the MFPTs as \cite{Ga09}:
\begin{eqnarray*}
T( x^{st}_l) & = & 2\int_{x^{st}_l}^{x^{un}_b}\frac{1}{\psi(y)}dy\int_{0}^{y}\frac{\psi(z)}{B(z)}dz, \; \mbox{and}\\
T(x^{st}_u) & = & 2\int_{x^{un}_b}^{x^{st}_u}\frac{1}{\psi(y)}dy\int_{y}^{\infty}\frac{\psi(z)}{B(z)}dz,\\
\end{eqnarray*}
where
\begin{equation*}
\psi(x)=\exp\left(\int_{x_{0}}^{x}\frac{2A(x')}{B(x')}dx'\right),
\end{equation*}
with $x_{0}=0$ for the $x^{st}_u \rightarrow x^{st}_l $ transition and
$x_{0}=x^{un}_b$ for the $x^{st}_l \rightarrow x^{st}_u $
transition.

The effect of changing $a$ and also the correlation parameter
$\lambda$ on the MFPT are shown in Fig.~\ref{MFPT} (for the details of
parameter values see the caption of Fig.~\ref{MFPT}).  We can observe
that the MFPT $T( x^{st}_l)$ decreases and $T( x^{st}_u)$ increases
with an increase in $a$ (see Fig.~\ref{MFPT}(a)) and the opposite
trend is observed for an increase in $\lambda$ (see
Fig.~\ref{MFPT}(b)).  These also imply the fact that an increase in
$a$ promotes the regime shift from left potential well to the right
potential well and vice versa for an increase in $\lambda$.  As MFPT
$T( x^{st}_l)$ approaches to zero at the bifurcation point, the system
loses its stability at low protein concentration.  The asymmetry in
the exit time is calculated by the ratio $\gamma= \frac{T(
  x^{st}_u)}{T( x^{st}_l)}$ and it diverges at the bifurcation point
where $T(x^{st}_l)$ approaches to zero (see Fig.~{\ref{MFPT1}})
\cite{Ga09}.  The quantity $\gamma$ thus gives an early warning signal
of regime shifts from low to high protein concentration state for an
increase in $a$.

\subsection{Early warning signals \label{sec5}}

\begin{figure*}[!ht]
\begin{center}
\begin{tabular}{lll}
{\footnotesize (a)} & {\footnotesize (b)} & {\footnotesize (c)}
\\ \resizebox{!}{0.83in}{\includegraphics{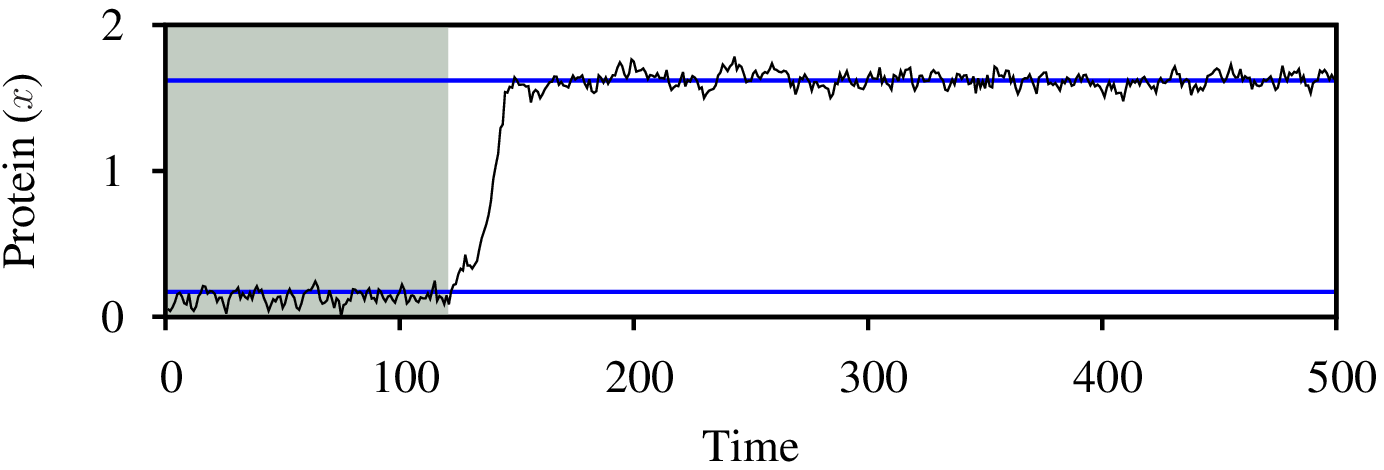}}\hspace{-0.08in}
& \resizebox{!}{0.82in}{\includegraphics{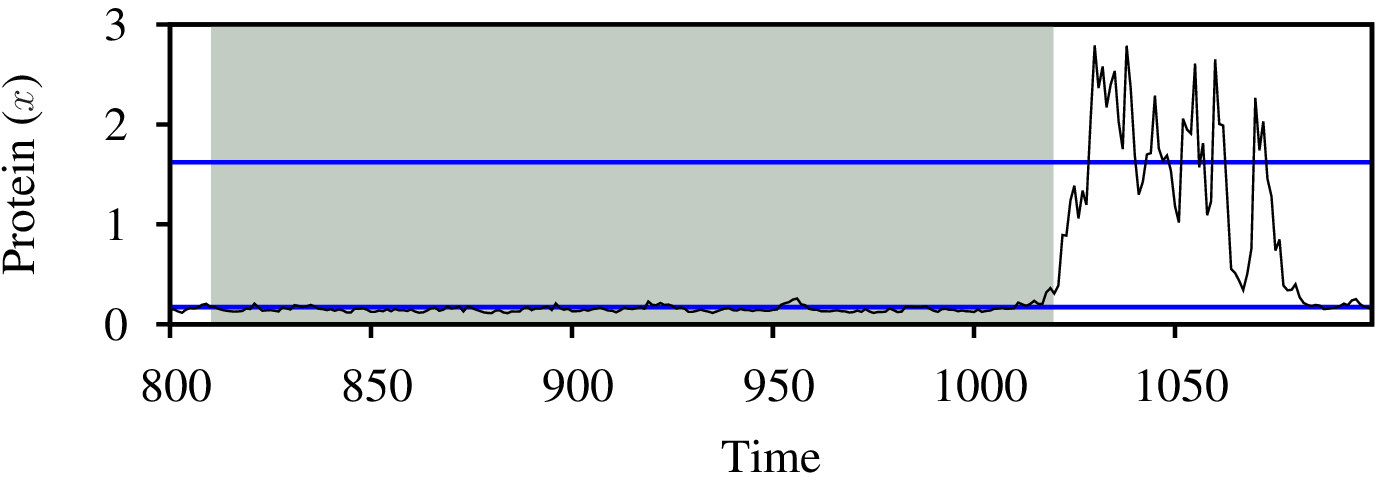}}
\hspace{-0.07in} &
\resizebox{!}{0.83in}{\includegraphics{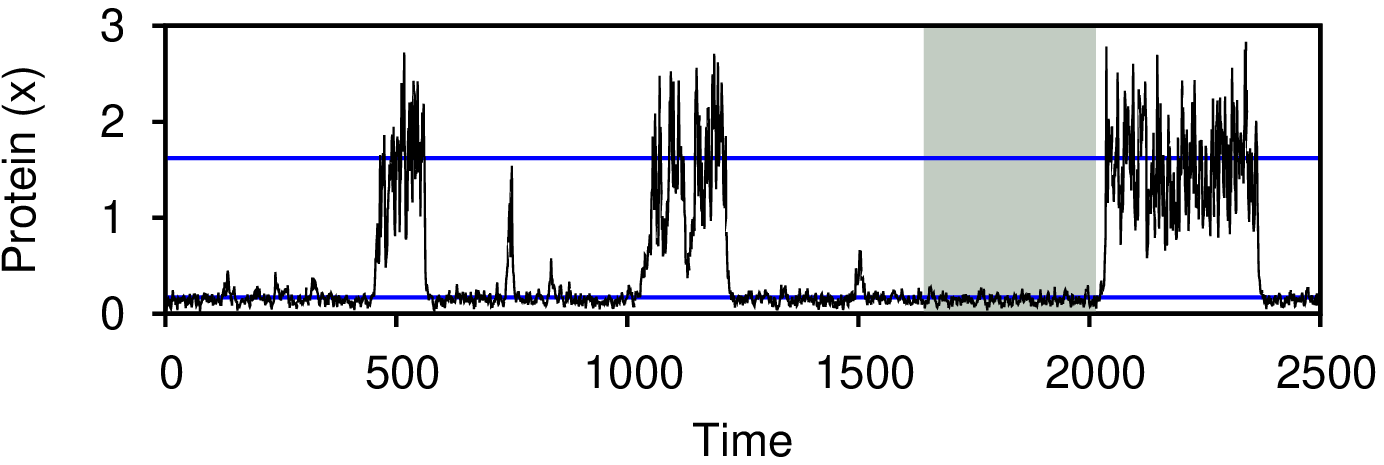}} \vspace{0.1in}
\\ \resizebox{!}{2.3in}{\includegraphics{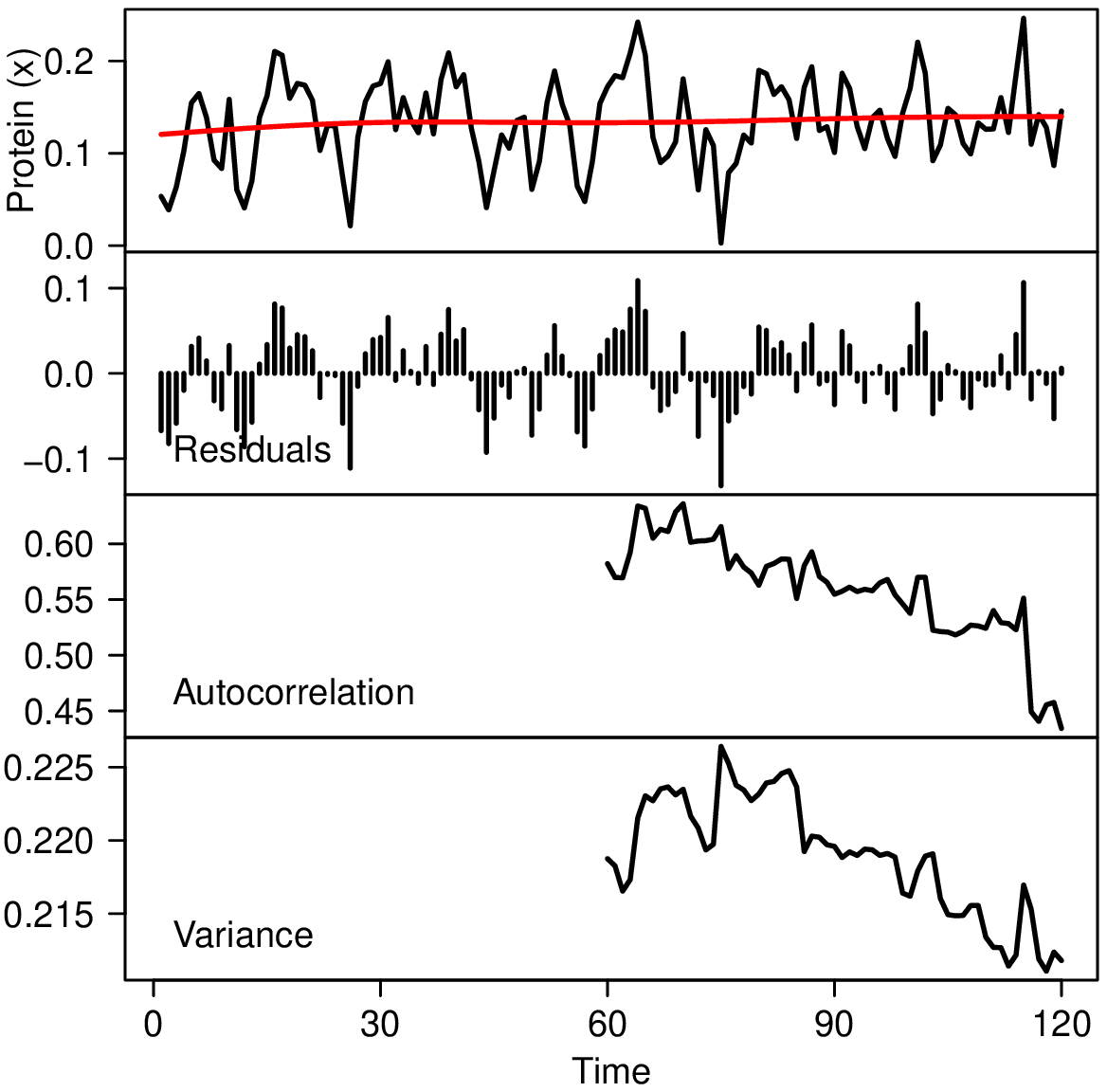}}\hspace{-1.in}
& \resizebox{!}{2.3in}{\includegraphics{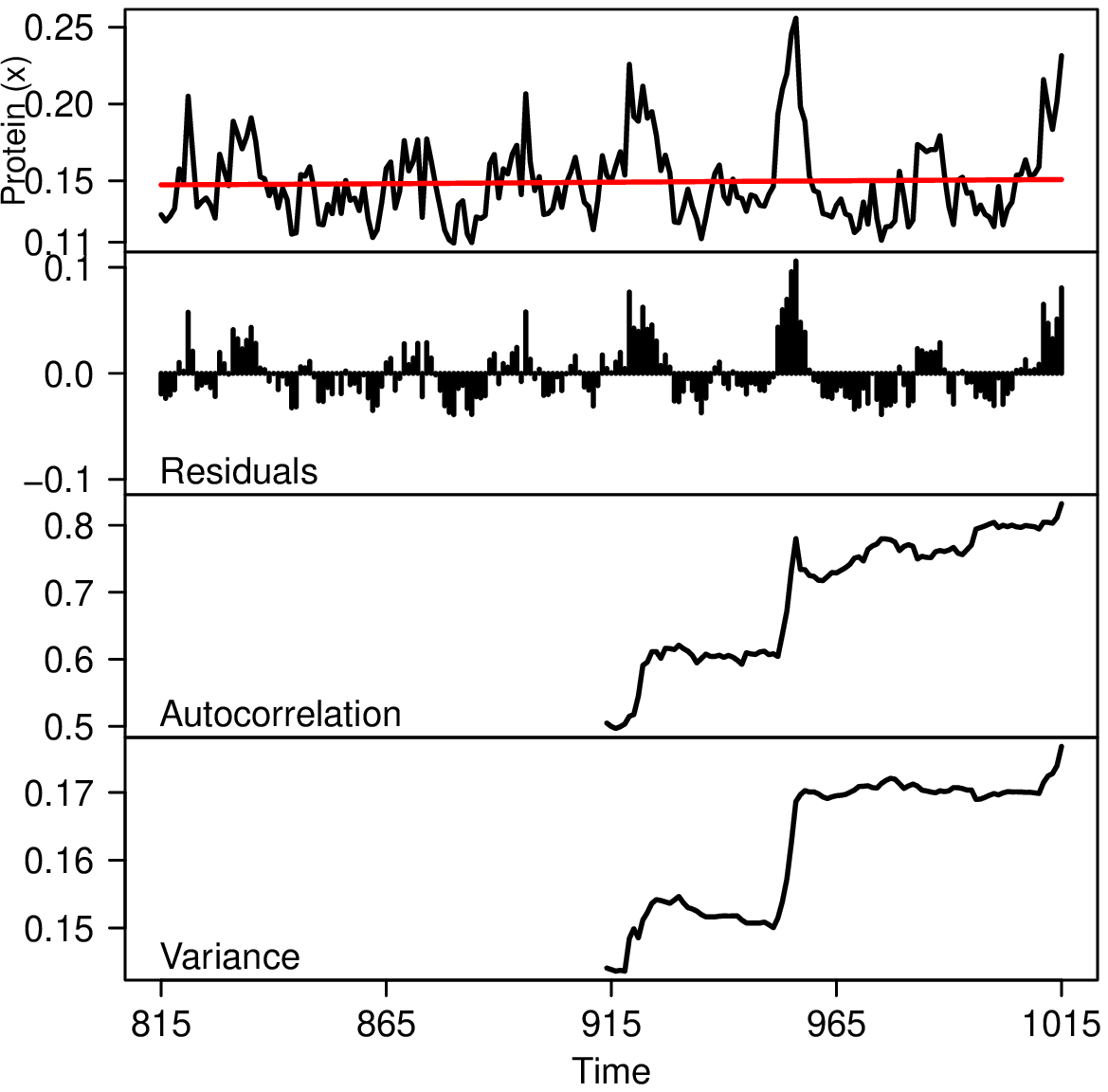}}
\hspace{-0.08in}  & \resizebox{!}{2.3in}{\includegraphics{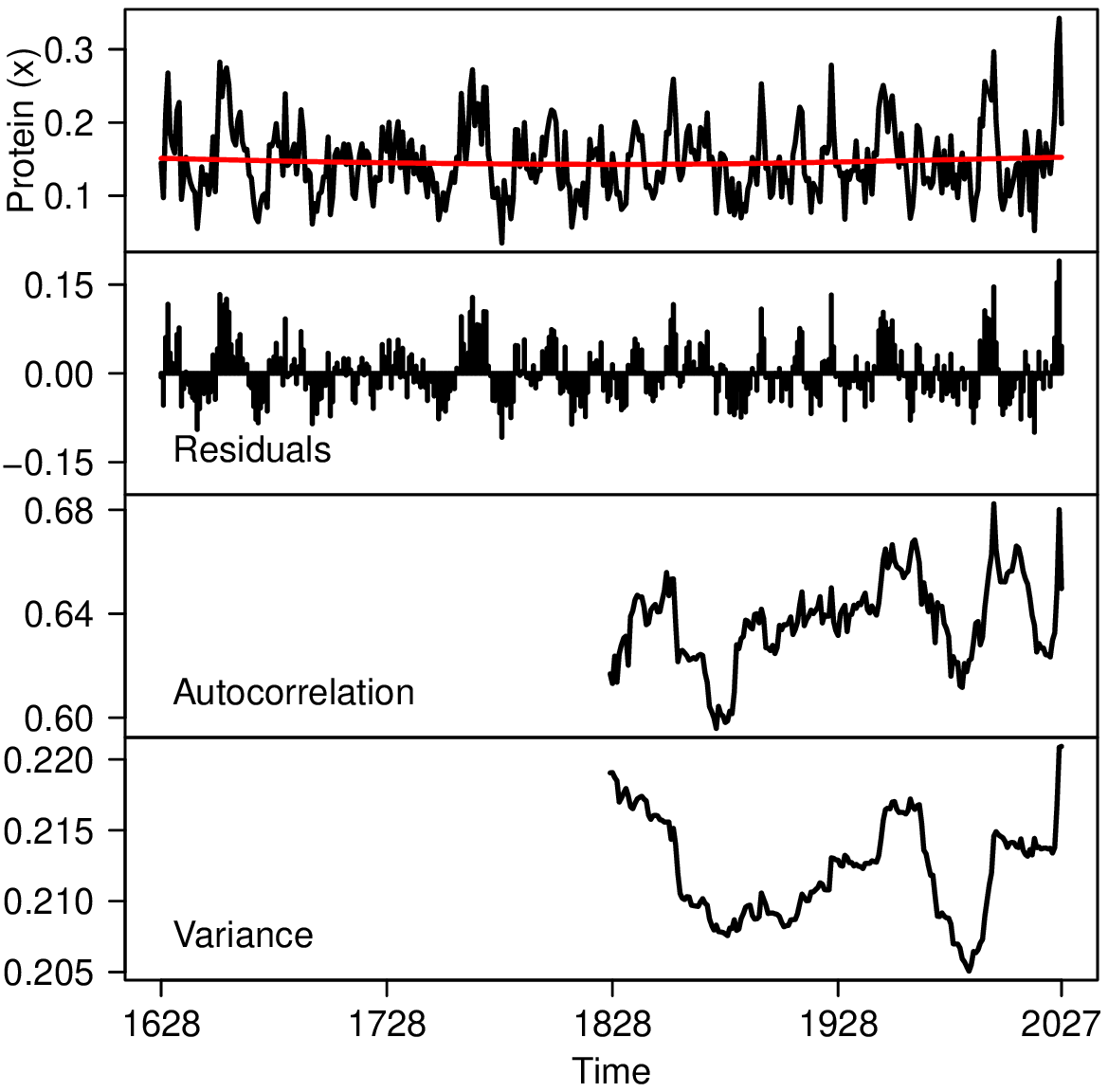}}  \\
\end{tabular}
\caption{\label{ss} Early warning signals for simulated time series
  data of the stochastic model in the case of stochastic switching.
  Here also we calculated variance and autocorrelation within rolling
  window of half the length of the time series segment (a segment is
  indicated by shaded region): (a) Additive noise: parameter values
  are $r=0.1$, $a=2.1$, $\sigma_{1}=0$ and $\sigma_{2}=0.9$; (b)
  Multiplicative noise: parameter values are $r=0.1$, $a=2.3$,
  $\sigma_{1}=0.9$ and $\sigma_{2}=0$ and (c) Correlated noise:
  parameter values are $r=0.1$, $a=2.1$, $\sigma_{1}=0.4$,
  $\sigma_{2}=0.65$ and $\lambda=0.5$.  See text for details.}
\end{center}
\end{figure*}

Here we assess the robustness of early warning signals to forewarn
upcoming shifts to an alternative regime by analyzing simulated time
series data \cite{Dakos:2012pone,Kefi:2014pone} from the considered
stochastic model of genetic regulation.  The simulation approach is
showing how the characteristics captured in our analytical
calculations can also be seen within individual time series
realizations based on specific parameter sets.  Early warning signals
are observable statistical signatures that antecede some state shifts.
These signals have mainly been derived from the phenomenon of critical
slowing down (CSD) which arises in the vicinity of bifurcation wherein
its dominant eigenvalue will cross zero \cite{Scheffer:2012sc}.  As
the eigenvalue reduces to zero, the response of the system becomes
very slow in recovering from perturbations and certain statistical
features, such as {\em increased variance} and {\em lag-$1$
  autocorrelation}, are predicted to appear in the time series
analysis as a result (reviewed in Scheffer et al. \cite{ScJo09}).
Although many regime shifts appear as sudden transitions to another
state in natural systems, the actual fact that not all regime shifts
are associated with a bifurcation \cite{Boettiger:2013br,DaCa15}.
Hence not all regime shifts associated with different mechanisms are
expected to exhibit CSD.  In Dakos et al. \cite{DaCa15}, it is pointed
out that CSD based early warning signals ``are not a panacea for
anticipating all types of regime shifts''.

Below we present two mechanisms those are responsible for observable
regime shifts in our simulated time series.  One is associated with
the saddle-node bifurcation and exhibits CSD and another is associated
with stochastic switching (SS) (i.e., purely noise-induced transition
and not associated with bifurcation) and does not exhibit CSD.  There
has been a recent debate about the success of early warning signals
for predicting stochastic switching induced regime shifts
\cite{Drake:2013cb,Boettiger:2013jc}.  Keeping that in mind, we felt
that it is worthwhile to present what occurs in our system.  For our
analyses, we consider three different stochastic time series, each for
CSD and SS: first we consider the presence of additive noise only,
second we consider the presence of multiplicative noise only and
finally presence of both noises with correlation in the system.  To
obtain the time series, stochastic simulations were performed in
MATLAB (R2011a) using the Euler-Maruyama method \cite{DH01} and a
standard integration step-size of 0.001.

In our simulated time series, we visually identify shifts between low
to high protein concentration and vice versa for both the cases, CSD
and SS (see Figs.~\ref{csd} and \ref{ss}).  Then we took different
time series segments (the gray shaded regions in Figs.~\ref{csd} and
\ref{ss}) of different lengths (keeping in mind that their time
lengths should be less than their MFPTs) preceding a regime shift and
analyzed those time series for the presence of early warning signals.
The ``Early Warning Signals Toolbox"
(http://www.early-warning-signals.org/) is used to perform the
statistical analyses.  To ensure stationarity in residuals, we used
Gaussian detrending with bandwidth 40, on the time series data before
performing any statistical analysis.  Then using a moving window size
of half the length of the considered time series (i.e., $50\%$ of the
considered time series segment), we calculate the variance and
autocorrelation in our state variable, $x$, as these two indicators
are both very commonly applied to anticipate regime shifts.  The
autocorrelation at lag-1 is measured by the autocorrelation function
(ACF): $$\rho_{1}=\frac{E[(x(t)-\mu)(x(t+1)-\mu)]}{\sigma^2},$$ where
$x(t)$ is the value of the state variable at time $t$, and $\mu$ and
$\sigma^2$ are the mean and variance of $x(t)$.  Variance is the
second moment around the mean $\mu$ and measured
as: $$\sigma^2=\frac{1}{N}\sum_{i=1}^{N}(x(t)-\mu)^2,$$ where $N$ is
the number of observations within the considered time frame.  A
concurrent increase in both of these indicators forewarn an impeding
regime shift \cite{Scheffer:2012sc}.

In Table~I, we summarize results of early warning signals (see
Figs.~\ref{csd} and \ref{ss}) for the considered two different cases
of regime shifts with three subcases each.  The signals [variance
  ($\sigma^2$) and autocorrelation ($\rho_1$)] are indicated as
``$+$'' if there is a concurrent rise; indicated as ``$+/-$'' if there
is a rise, but some false pick in the signals before the regime shift
due to the presence of large fluctuations in the data; and indicated
as ``$-$'' if looking at the signals, it is not possible to forewarn an
impending regime shift.
\begin{table}[!ht] 
\label{t1}
\centering
\caption{Two mechanisms, CSD and SS that can produce regime shifts
  with and without bifurcation in the system and the outcome of early
  warning signals for regime shifts.}
\begin{tabular}{l l l l l}
\hline
\multicolumn{5}{c}{\done \vspace{-.1in}}\\
\multicolumn{5}{c}{\done \hspace{0.3in}  \hspace{0.45in} CSD  \hspace{0.85in} SS}\\
\multicolumn{5}{c}{\done \vspace{-.1in}}\\
\multicolumn{5}{c}{\done \hspace{0.05in} Noise Type  \hspace{0.15in} Var($\sigma^2$) 
\hspace{0.01in} ACF($\rho_1$) \hspace{0.01in} Var($\sigma^2$) \hspace{0.01in} ACF($\rho_1$)}\\
\hline
\vspace{-.1in}
\high &&&&\\
\high $~~$ {\em Additive} & $~~~~~~$+ & $~~~~~~~~$+ & $~~~~~~$$-$ & $~~~~~~~~~$$-$\\
\vspace{-.12in}
\high &&&&\\
\hline
\vspace{-.1in}
\low &&&&\\
\low $~${\em Multiplicative}$~$ & $~~~~~$ + & $~~~~~~~$ +/$-$ & $~~~~~~$+ & $~~~~~~~~~$+\\
\vspace{-.12in}
\low &&&&\\
\hline
\vspace{-.1in}
\high &&&&\\
\high $~$ {\em Correlated: }  &  &  & & \\ 
\high $~$ {\em Additive and}  & $~~~~~$ + & $~~~~~~~$ +/$-$ & $~~~~~$ $-$ & $~~~~~~~~~$$-$\\
\high $~$ {\em multiplicative$~~$}  &  &  & & \\
\hline
\end{tabular}
\end{table}
The result in Table~I shows that in the case of CSD with all three
different types of stochasticity applied to model (\ref{Eq:main}), the
variance is robust and always successfully predict upcoming regime
shifts for all the three types of noise.  However, the lag-1
autocorrelation is only successful to predict regime shifts for the
case of additive noise only and can't predict very positively an
upcoming regime shift for multiplicative and correlated noise.  This
establishes the fact that even for the case of CSD the variance is
more robust than autocorrelation in detecting regime shifts in genetic
regulation and it is line with the findings in \cite{Da12}.  For SS,
both of the early warning signals are not very successful in detecting
regime shifts.  Using the EWS indicators it is not possible to detect
upcoming regime shifts for additive and correlated noises (see
Fig.~\ref{ss}).  However, in the presence of multiplicative noise, the
early warning signals give positive results.  This is indeed a nice
result given the fact that early warning signals are not developed
for noise induced transitions, but rather for CSD which is associated
with bifurcations.  Nevertheless, our results show that even for the
case of CSD though the EWS somewhat gives positive results, still some
regime shifts may not be triggered by {\em a concurrent} rise in
variance and autocorrelation.  This is due to the fact these CSD
indicators are reliable for specific systems and also have some key
limitations \cite{DaCa15}.  Like, there is false positive and false
negative errors in data \cite{SKP14}.  Some other limitations include
size of the window which varies across the literature and how much
data are required for analyzing EWS is still not clear.  Moreover,
variance is a robust indicator as compared to autocorrelation in the
case of CSD and this is due to the fact that autocorrelation is
affected by the length of time series (see
Appendix S2 for more details) and more sensitive to false
alarms.  In the case of stochastic switching, anticipating regime
shifts is more difficult due to sudden transitions.  The significance
of these findings are discussed in the discussion section below.

\section{\label{con} Discussion}

In this paper, we investigated a stochastic genetic regulatory circuit
using analytical techniques and numerical simulation to anticipate
regime shifts in protein concentration.  In this respect, we derived
an approximate Fokker-Planck equation from the Langevin equation.  We
studied the effects of intensity of both the additive and
multiplicative external noise and their correlation on the gene
regulatory system.  The present study suggests that the presence of
additive noise in the system induces regime shifts from a low protein
concentration to a high protein concentration state, whereas
multiplicative noise induces regime shifts from a high to a low
protein concentration state.  Moreover, we also show that how a
correlation between the additive and multiplicative noise is important
in determining regime shifts and hence can regulate the production of
protein levels.  Furthermore, an increase in the cross correlation
intensity from a negative to a positive value between the two noises
induce regime shifts from high to low protein concentration state.
Here we have also computed the robustness of steady states using the
MFPT \cite{Ga09}.  Our MFPT result uncovers the fact that an increase
in MFPT of right potential well with the maximum production rate
promotes regime shifts from left potential well to the right potential
well and vice versa for an increase in the cross correlation
intensity.

In addition, we used a recently developed tool
(http://www.early-warning-signals.org/) of early warning signals to
anticipate regime shifts in the considered gene regulatory system
using simulated time series data for both the CSD and SS.  Extensive
literature on regime shifts is available for ecosystems, financial
markets, climatic shifts, etc.  However, to the best of our knowledge,
there is very less work available on anticipating regime shifts in
many areas of developmental biology, specifically in genetic
regulation \cite{ChLi12,Ya14}.  Here for the first time, for a genetic
regulatory system, we used the time series analysis based EWS approach
to predict upcoming regime shifts.  As anticipated by some previous
authors \cite{Drake:2013cb,Boettiger:2013jc}, we also find that the
EWS of raising variance and autocorrelation are in general sensitive
to false alarms and not always successful to reliably predict
impending state shifts in our model.  We found that the EWS are
moderately more successful when we analyzed time series data in
advance to a state shift in the case of CSD, whereas in the case of
SS, it is specific to particular noise only.  We observed some key
limitations and statistical issues of EWS, such as false pick in data,
size of the window and length of time series data.  For our model, we
also verified that rising variance appears as a robust indicator of
CSD as compared to lag-$1$ autocorrelation.  This is mainly due to the
fact that autocorrelation is sensitive to the length of time series.
For accurate estimation of autocorrelation, we need long and
equidistant time series data which are not always available for real
systems.  However, variance is insensitive to this effect and as a
result, measuring autocorrelation as EWS may increase the possibility
of false alarms.  While testing EWS to our simulated time series data,
we also observed that selection of window size is another major
problem for getting a positive signal of regime shifts.  In the case
of SS, anticipation of regime shifts is extremely difficult because a
bifurcation point is neither approached nor crossed and there is
suddenly a phase transition.

Anticipating regime shifts in gene regulatory system (aka in protein
concentration level) can be useful to prevent disease onset and
progression which may intercept unacceptable sudden transitions from a
healthy state to a disease state \cite{TrAn15}.  Examples of such
regime shifts are asthma attacks, epileptic seizures and sudden
deterioration of complex diseases \cite{ChLi12,gl15,TrAn15}.  A well
documented example of regime shift is type 1 diabetes (T1D) which is a
form of diabetes mellitus \cite{LiLiZh2013}. T1D is a chronic
inflammatory disease caused by insufficient production of insulin by
$\beta$ cells in the pancreas.  The genetic association of T1D is that
the production of insulin through $\beta$ cells depends on
HLA-encoding genes \cite{NoEr12}.  If T1D associated genes are in
``on'' expression state, then $\beta$ cells produce insulin and
releases insulin into the blood stream, but if they are in ``off''
expression state $\beta$ cells fail to produce insulin which leads to
T1D.  Prediction of regime shifts in gene expression using EWS from
normal (``on'') to diabetic (``off'') state could prevent the switch
to diabetic state and help to maintain the level of insulin.

In summary, our work reveals that stochasticity can have diverse
complex effects on genetic regulatory systems.  Early warning signals
to anticipate forthcoming regime shifts in gene expression requires
special attention to the underlying various statistical issues and
limitations.  In addition, to select a suitable window size and data
length raise further difficulties.  The main challenge of detecting
early warning signals includes risk of false alarms and failed
detections.  One needs a deeper understanding of early warning signals
of regime shifts, and how a balance between early warning signals and
false alarms is achieved, will lead to important new insights in
genetic regulation.  Moreover, our results establish the important
fact that finding a more robust indicator of regime shifts in complex
natural systems is still in its infancy and demands extensive
research.  We hope that this study may also increase the interest
among researchers to find a more robust indicator for detecting
upcoming sudden transition in much broader class of systems in
developmental biology.

\section*{Supporting Information}

\noindent
{\bf Appendix S1} $~$ {\bf Derivation of Stationary Probability
  Density Function.} $~$ (PDF)\\

\noindent
{\bf Appendix S2} $~~$ {\bf Additional examples of early warning signals.} $~$ (PDF)
\section*{Acknowledgments}
P.S.D. acknowledges financial support from ISIRD, IIT Ropar Grant No.
IITRPR/Acad./52. The authors thank T. Banerjee and Ramesh A. for their
helpful comments on the manuscript. We thank David Frigola for sharing
his code on the stationary probability distribution.

\section*{Author Contributions}
Conceived and designed the experiments: YS$~$  PSD. Performed the
experiments: YS$~$  PSD. Analyzed the data: YS$~$  PSD. Contributed
reagents/materials/analysis tools: YS$~$  PSD$~$  AKG. Wrote the paper: PSD$~$  YS.


\end{document}